# The Arctic Ocean seasonal cycles of heat and freshwater fluxes: observation-based inverse estimates

Takamasa Tsubouchi[*], Sheldon Bacon and Yevgeny Aksenov

National Oceanography Centre, Southampton, UK

Alberto C. Naveira Garabato

University of Southampton, Southampton, UK

Agnieszka Beszczynska-Möller

Institute of Oceanology PAS, Physical Oceanography Department, Sopot, Poland

Edmond Hansen

Norwegian Polar Institute, Tromsø, Norway and Multiconsult, Tromsø, Norway

Laura de Steur

Norwegian Polar Institute, Tromsø, Norway

Beth Curry and Craig M. Lee

University of Washington, Seattle, Washington, the USA

---

[*] Corresponding author address: Geophysical Institute, University of Bergen and Bjerknes Centre for Climate Research, Bergen, Norway; E-mail: takamasa.tsubouchi@uib.no

ABSTRACT

This paper presents the first estimate of the seasonal cycle of ocean and sea ice net heat and freshwater (FW) fluxes around the boundary of the Arctic Ocean. The ocean transports are estimated primarily using 138 moored instruments deployed in September 2005 – August 2006 across the four main Arctic gateways: Davis, Fram and Bering Straits, and the Barents Sea Opening (BSO). Sea ice transports are estimated from a sea ice assimilation product. Monthly velocity fields are calculated with a box inverse model that enforces volume and salinity conservation. The resulting net ocean and sea ice heat and FW fluxes (annual mean ± 1 standard deviation) are 175±48 TW and 204±85 mSv (respectively; 1 Sv = $10^6$ $m^3$ $s^{-1}$). These boundary fluxes accurately represent the annual means of the relevant surface fluxes. Oceanic net heat transport variability is driven by temperature variability in upper part of the water column and by volume transport variability in the Atlantic Water layer. Oceanic net FW transport variability is dominated by Bering Strait velocity variability. The net water mass transformation in the Arctic entails a freshening and cooling of inflowing waters by 0.62±0.23 in salinity and 3.74±0.76°C in temperature, respectively, and a reduction in density by 0.23±0.20 kg $m^{-3}$. The volume transport into the Arctic of waters associated with this water mass transformation is 11.3±1.2 Sv, and the export is -11.4±1.1 Sv. The boundary heat and FW fluxes provide a benchmark data set for the validation of numerical models and atmospheric re-analysis products.

# 1. Introduction

The Arctic has experienced unprecedented climate change in last few decades. Surface air temperature in the Arctic has risen more than twice as fast as the rest of the globe, a phenomenon known as "Arctic amplification" [Serreze et al. 2009]. The September sea ice extent has steadily decreased since 1979 [Serreze et al. 2007a; Serreze et al., 2016]. The Greenland ice cap is melting at an increasing rate [Velicogna, 2009; Shepherd et al. 2012]. Russian river runoff has increased [Overeem and Syvitski 2010; Shiklomanov and Lammers, 2014]. Atlantic water (AW) temperature shows a warming signal in 2000's in Fram Strait [Beszczynska-Möller et al. 2012], offshore of the Laptev sea [Polyakov et al. 2011], and in the central Arctic [Korhonen et al. 2013; Polyakov et al. 2012]. Freshwater (FW) storage in the Canadian basin increased significantly during the 2000s [Proshutinsky et al., 2009; Giles et al., 2012; Rabe et al., 2014; Armitage et al., 2016]. This steepened the sea surface height gradient, driving an acceleration of the Beaufort gyre circulation during the period [McPhee et al. 2009; Giles et al. 2012; McPhee 2013; Armitage et al., 2017]. Polyakov et al. [2017] identify what they call the encroaching "atlantification" of the Eurasian Basin as a developing process that is proelling the Arctic towards a new climate state, and they associate sea ice reductions, halocline weakening, shoaling of the AW layer, and increased winter ventilation with this process.

Our understanding of the role of the ocean in the Arctic climate system is improving, but our knowledge of the fundamental physical fluxes of heat and freshwater (FW) remains poor [Mauritzen et al., 2011; Haine et al., 2015; Carmack et al., 2015, 2016]. Even the long-term mean oceanic heat transport across the Arctic boundary is unknown [Schauer and Beszczynska-Möller 2009; Carmack et al., 2015]. For the oceanic FW transport, decadal average views based on different available data periods represent the state of the art [Dickson et al., 2007; Haine et al., 2015; Carmack et al., 2016]. This basic deficiency is well illustrated by Cowtan and Way [2014]. In their Fig. 1, they show four different estimates of Arctic surface temperature trends including three near global reconstruction products: extrapolated surface temperature product, satellite surface temperature product, re-analysis product. Their Fig. 1a highlights starkly the sparsity of sustained surface temperature measurements over the Arctic Ocean. This sparsity applies to all other climate-relevant parameters over the Arctic Ocean, such as wind speed and direction, humidity, evaporation, precipitation, etc. A parallel problem exists with river runoff: the largest rivers are gauged, but ~30% of total runoff is estimated to be ungauged [Lammers et al. 2007; Shiklomanov and Lammers, 2009].

To begin to go some way towards filling this knowledge gap, we have been developing the "control volume" approach to the calculation of Arctic ice and ocean fluxes of climate-relevant quantities. A control volume is a box, which, in this case, comprises the Arctic Ocean, where the upper surface is either the surface of the open ocean or the upper surface of sea ice (as appropriate), the lower surface is the sea bed, and the sides comprise either land or ocean passages ("gateways"). With a control volume, simple conservation laws can be applied, using inverse methods, to measurements around the ice and ocean boundary of the box, in order to ensure conservation of mass and of salinity. The output is a mass- and salinity-conserving ice and ocean boundary velocity field. This field then enables the calculation of surface fluxes of heat and FW, allowing for storage and release of those quantities within the control volume interior. The waters entering the Arctic are mainly warm and salty; those leaving are mainly cold and fresh; and the net change in properties is caused by surface fluxes.

Quasi-synoptic ocean and sea ice heat and FW fluxes were calculated across the pan-Arctic boundary for summer 2005 by Tsubouchi et al. [2012; hereafter T2012]. T2012 was the first paper to demonstrate the power of the control volume approach, by assembling measurements across the four main Arctic gateways – Fram, Bering and Davis Straits, and the Barents Sea Opening (BSO). T2012's boundary data included high-resolution quasi-synoptic hydrographic measurements (of temperature, salinity, velocity and pressure), supplemented by sea ice export values based on upward-looking sonar measurements of sea ice thickness and satellite measurements of sea ice area flux, and by a high-resolution ice-ocean general circulation model, to provide "patching" in small regions where data were lacking. Subsequent analysis used the T2012 boundary velocity field to generate baseline estimates of pan-Arctic fluxes of inorganic and organic nutrients [Torres Valdes et al. 2013, 2016] and carbon [MacGilchrist et al. 2014].

Here we take a similar approach, but our aim now is to calculate the first annual average values of Arctic ice and ocean boundary – and hence also of surface – heat and FW fluxes. The approach is made possible by the sustained presence in the four main Arctic gateways – Fram, Bering and Davis Straits, and the BSO – of moored instrumentation measuring temperature, salinity and velocity. In combination, as an Arctic Ocean Boundary Array, they represent an under-utilised resource for the estimation of Arctic surface fluxes. The objectives of this study are, therefore, (i) to generate for the first time a time series of twelve, monthly, piecewise-continuous estimates of the boundary components of net (surface) fluxes of heat and freshwater, in order to expose the seasonality inherent in that component, and also (ii) to examine the extent to which the assessment quality is dependent on the quantity and distribution of the available measurements.

This paper is structured as follows. We present our data and methods in section 2, our results in section 3, and in section 4 we provide a discussion and interpretation of the results. Section 5 is a summary.

## 2. Data and methods

In this section, we first describe our measurement and model output resources; then we describe the method by which a year-long time series of gridded property and velocity fields is generated; third, we set out a brief overview of the box inverse model; finally, we describe the method of heat and FW flux calculation.

*a. Hydrographic data and numerical model outputs*

Our main data source is 138 moored instruments deployed in the Arctic four main gateways from summer 2005 to summer 2006. Moored instrument locations (41 mooring sites) and the positions in the water column of the 138 instruments are shown in Fig. 1. Instruments comprise temperature- and salinity-measuring devices, acoustic Doppler current profilers (ADCP), and single-point current meters. Table 1 summarises the moored instruments analyzed in this study. In the gateways, the following moored data are analyzed: 39 moored instruments (21 MicroCATs, 9 RCMs and 9 ADCPs) in Davis Strait from 1 September 2005 to 30 September 2006 [Curry et al. 2014]; 17 moored instruments (5 MicroCATs and 12 RCMs) in the western part of Fram Strait from 5 September 2005 to 3 September 2006 [Hansen et al. 2006; de Steur et al. 2014]; 60 moored instruments (16 MicroCATs and 44 RCMs) in the central and eastern parts of Fram Strait from 26 August 2005 to 26 August 2006 [Fahrbach and Lemke 2005; Beszczynska-Möller et al. 2012]. In the BSO, 13 moored instruments (11 RCMs and 2 ADCPs) from 18 August 2004 to 29 June 2007 [Ingvaldsen et al. 2004]; 4 moored instruments (3 MicroCATs and 1 RCM) in the Russian side of Bering Strait from 20 August 2005 to 24 August 2006 [R/V Sever Cruise Report 2005]; 5 moored instruments (3 MicroCATs and 2 RCMs) in the US side of Bering Strait from 11 July 2005 to 31 August 2006 [Woodgate 2006]. In Davis Strait, processed daily data is analyzed instead of data at the original sampling intervals [Curry et al., 2014]. A total of 138 moored instruments records (46 MicroCATs, 81 RCMs and 11 ADCPs) are combined between 5 September 2005 and 24 August 2006.

Temperature and conductivity (for salinity) are measured by SeaBird Electronics MicroCATs (SBE16, SBE37). The SBE16 initial accuracy is 0.005°C for temperature and 0.0005 S/m for conductivity. The SBE37 initial accuracy is 0.002°C for temperature and 0.0003 S/m for conductivity. Pressures are measured in some cases. Single point velocities (direction and speed) are measured by Aanderaa RCMs (Rotor Current Meter: RCM7 and RCM8; Doppler Current Meter: RCM9 and RCM11). The RCM7/8 initial accuracy is 5-7.5° for direction and 0.01 ms$^{-1}$ or 4% of recorded value for speed. The RCM9/11 initial accuracy is 5-7.5° for direction and 0.0045-0.0050 ms$^{-1}$ as statistical precision (standard deviation) for speed. All Aanderaa RCMs measure temperature, some also measure conductivity. RCM temperature and conductivity accuracies are 0.05°C and 0.0015 S/m, respectively. The time resolution of the moored instruments ranges from 15 min to 180 min.

Nine CTD sections occupied nearly monthly between Bear Island and Norway in BSO are also analyzed. For each section, mean station spacing is 30 km with 1 dbar vertical resolution. A typical section comprises 20 stations. The hydrographic data are downloaded from the International Council for the Exploration of the Sea database (http://www.ices.dk/).

The Nucleus for European Modelling of the Ocean (NEMO) is a widely-used framework for oceanographic modelling. Bacon et al. [2015] describe the model configuration and forcing. The model's tripolar grid concentrates resolution in the Arctic, so that the 1/12° configuration used here has effective horizontal resolution around the defined boundary of ~3 km. Model output is available as 5-day mean fields of ocean temperature, salinity and velocity. It has 75 layers in the vertical, of which there are 24 layers in the upper 100 m, 11 layers between 100–300 m, and 25 layers between 300–3000 m. There are no deeper layers around the Arctic boundary. We use model output to patch in data in regions where observations are lacking (Belgica Bank, Bjørnoya Bank, and the water above the shallowest instruments). The model output is available at http://gws-access.ceda.ac.uk/public/nemo/. The model's credibility in the Arctic context has been established in a number of studies, for example: Atlantic water inflow [Aksenov et al. 2010a; Lique and Steele, 2012, 2013], Polar water outflow [Aksenov et al. 2010b; Lique et al., 2010], Arctic boundary current [Aksenov et al. 2011] and the East Greenland Coastal Current [Bacon et al. 2014].

Monthly sea ice thickness, velocity and temperature are obtained from the Pan-Arctic Ice Ocean Modeling and Assimilation System (PIOMAS; Zhang and Rothrock, 2003). Schweiger et al. [2011] estimate PIOMAS sea ice thickness uncertainty to be 0.2 m. The sea ice thickness (m), sea ice velocity ($ms^{-1}$) and surface temperature (˚C) data are downloaded from the Polar Science Centre at UW (http://psc.apl.uw.edu/research/projects/arctic-sea-ice-volume-anomaly/). Note that ice concentration is included in the calculation of sea ice thickness at a given point by PIOMAS output.

The vertical ocean boundary area is 1,051 $km^2$, and the quasi-horizontal surface area inside the boundary is $1.13 \times 10^{13}$ $m^2$ [Jakobsson et al., 2002]. The mooring and CTD measurements span a vertical ocean area of 972 $km^2$, or 92% of the total. The area patched by model output is 79 $km^2$, 8% of the total.

*b. Construction of 5 days temperature, salinity and velocity gridded sections*

We choose to project the mooring records and CTD measurements in BSO onto the same temporal resolution as NEMO. Therefore we first need to establish a common (5-day) time base for the moored measurements. Our common time series starts at 12:00 on 30 August 2005 and ends at 12:00 on 04 September 2006.

With all instrumental records on a continuous 5 day time base, initial coast-to-coast and sea-bed-to-surface fields of temperature, salinity and cross-section velocity are created. The horizontal grid resolution is 3km, considering the 1/12 NEMO's horizontal resolution. The vertical grid resolution is 75 layers, as same as NEMO's. We add variability information above the shallowest moored instruments and across shelf regions using NEMO output. The detail mooring data availability and treatment, and gridding procedure can be found in online supporting information A.

*c. Box inverse model*

A box inverse model is applied to the Arctic boundary measurements to obtain velocity fields that conserve mass and salt. The 5-day mean fields within corresponding calendar months are averaged to obtain 12 monthly fields. In reality, 5-day mean fields are converted into 11 × 30-day (6 × 5-day time step) months plus 1 × 35-day (7 × 5-day time step) month, equivalent to July 2006. The inverse model method is based on that used by T2012, where full details are found. Settings of the box inverse model used in this study is provided in Appendix A.

These are four differences between this study and T2012. First, we use a sea ice data assimilation product (PIOMAS) to obtain an observation-based, continuous estimate of sea ice exports across the four main gateways. Second, the use of NEMO model output is more extensive because the moored measurements reach neither the coasts nor the sea surface. In particular, two wide banks lack measurements – Belgica Bank, off north-east Greenland, and Bjørnoya Bank, between Bear Island and Svalbard; and the shallowest instruments lie (for practical reasons) between 50–100 m water depth. Third, T2012 was a quasi-synoptic study using measurements made from ships (albeit within the same period of about a month) whereas this study is synoptic, employing a calendar year of continuous, simultaneous marine measurements. Fourth, T2012 used measurements of high spatial resolution, whereas here, the spatial (vertical and horizontal) resolution is determined by the moored instrument locations.

*d. Heat and FW transport calculation and uncertainties*

In this section we present the equations that determine the relationships between the surface fluxes of heat and FW and those fluxes at the ice and ocean boundary, including storage inside the control volume, as presented in Bacon et al. [2015]. However, in this study, we do not calculate the storage fluxes explicitly. For detailed consideration of the meaning of our results in this context, see section 4a below. For heat fluxes,

$$F_H^{surf} = - \iint \left[ \left(v'\rho c_p \theta'\right)^o + \left(v'\rho c_p \theta'\right)^i + (v\rho L)^i \right] ds dz + F_H^{stor} \quad (1)$$

where $F_H^{surf}$ is the surface heat flux and $F_H^{stor}$ is heat storage change in time within the boundary; superscripts *o* and *i* refer to liquid ocean and sea ice, respectively; $\rho$ is density; $c_p$ and $L$ are specific heat capacity and latent heat of freezing, respectively; $v$ and $v'$ are cross-section velocity and its anomaly about the mean, respectively; $\theta'$ is the potential temperature difference from its reference value; $z$ is depth and *s* the along-boundary coordinate. The reference value for potential temperature is the area-weighted section mean for each month. Oceanic temperature transport and sea ice sensible heat transport are calculated with reference to the appropriate boundary-mean potential temperature for each month. The annual mean (±1 sd) is 1.01±0.16 ˚C, and the 12 monthly values are 1.26, 1.15, 1.01, 0.91, 0.94, 0.91, 0.83, 0.82, 0.88, 0.99, 1.15, 1.24 ºC for September 2005 – August 2006. Annual mean (±1 sd) sea ice temperature is -11.6±8.9 ˚C, with maximum -0.54 ˚C in July 2006 and minimum -25.29 in February 2006, obtained from PIOMAS monthly output over the Belgica Bank region (78-80˚N, 17-6˚W). Sea ice specific heat capacity is obtained from Ono [1967]; sea ice density is fixed as 930 kg m$^{-3}$.

In similar manner, the surface FW flux is balanced with boundary ocean and sea ice FW transports and storage term in FW flux equation.

$$F_{vol}^{surf} = \frac{\iint [(v\prime S\prime)^o + (v\prime S\prime)^i] ds dz}{\bar{S}} + F_{FW}^{stor} \qquad (2)$$

where $F_{vol}^{surf}$ is the surface FW flux and $F_{FW}^{stor}$ is FW storage change in time within the boundary. $S\prime$ is the salinity difference from its reference value. The area-weighted mean salinity ($\bar{S}$) across the section is 34.67±0.02 (annual mean ±1 sd), and the 12 monthly values are 34.67, 34.64, 34.63, 34.63, 34.66, 36.67, 34.68, 34.68, 34.70, 34.68, 34.67, 34.66 during September 2005 – August 2006. Sea ice salinity is set as 6 throughout the year because PIOMAS does not calculate sea ice salinity. The sea ice salinity 6 is close to first year sea ice salinity [Kovacs et al., 1996; Shokr and Sinha, 2015].

We make a practical distinction between oceanic heat transport and oceanic temperature transport. We refer to a quantity as temperature transport when it is sensitive to choice of reference temperature. We refer to a quantity as heat transport when it is not sensitive to choice of reference temperature. The sensitivities of these quantities are illustrated in T2012 (Fig. 20). Following Talley [2003], the two quantities are distinguished (for clarity) through their units – Watts (as usual) for heat transport, and we introduce the Watt-equivalent, or "W-eq." for temperature transport. By analogy, oceanic FW transports are also distinguished depending on their sensitivity to choice of reference salinity through their units – "mSv" and "mSv-eq", following Talley [2008].

Finally, we present two different types of uncertainty. One is the temporal variability of a quantity over a given period, presented as one standard deviation (sd) of the relevant time series. The other is uncertainty as quantification by sensitivity tests or other estimation method. In order to distinguish two different types explicitly, we use italic when quoting "sensitivity" uncertainties and plain font for sd.

## 3. Results

In this section, we first illustrate the five-day mean fields of temperature, salinity and velocity in terms of annual mean values and variability at a selected depth. Next we describe the impact of the inverse model on the initial velocity fields, and third, we consider the resulting velocity fields and their horizontal and diapycnal volume transports. Fourth are temperature and heat transports, and fifth are FW transports, both of which include a useful decomposition to identify the major sources of variability. Sixth, we present the results of a suite of sensitivity studies directed at understanding the impact of the NEMO model output on the results. Finally, we consider the transport-weighted mean temperature, salinity and density of inflows and outflows and the associated net water mass transformation.

### a. *Five-day mean gridded sections*

Fig. 2a shows temperature variability at 50 m depth. Regarding the seasonal cycle, substantial variability is observed in Bering Strait (–2–8˚C), the eastern side of Davis Strait (–2–6˚C), the east of Fram Strait (4–7˚C), and south of Bear Island in the BSO (5–9˚C); these are all inflow regions. In contrast, in outflow regions (western Davis Strait, western Fram Strait including Belgica Bank), the temperature remains low throughout the year. Fig. 2b shows salinity variability at 50 m depth. It is notable that relatively fresh regions show highest seasonal variability: the outflows in Davis Strait and in the west of Fram Strait, and also the Bering Strait inflow. AW inflow salinities are relatively stable through the year.

Fig. 2c shows cross-section velocity at 50 m depth. Annual averaged values capture the major currents around the boundary. In Davis Strait, the outflow in the west and inflow in the east. In Fram Strait, the northwards current near the Greenland coast, the anticyclonic circulation over Belgica Bank, the EGC outflow and WSC inflow. In the BSO, there is a variable outflow just south of Bear island, and inflows in the main portion of the BSO, including Norwegian Coastal Current. The Bering Strait mean inflow (0.15 ms$^{-1}$) has large variability superimposed (±0.3 ms$^{-1}$). It is notable that Bering Strait exports seawater from the Arctic to the Pacific Ocean during November-December 2005; c.f. Woodgate et al. [2005].

*b. Inverse model modification*

Fig. 3 shows initial net volume and FW transport imbalances by month. The annual means of net (including surface) volume and FW transports are close to zero, at 0.8±3.1 Sv and 33±87 mSv (respectively). However, monthly imbalances are large. For example, November 2015 shows a volume transport deficit of 5.6 Sv and a FW transport surplus of 178 mSv. These imbalances are resolved applying volume and salt constraints using the box inverse model for consective twelve months.

For each of the twelve months, the row- and column weighted system of equations is solved by singular value decomposition [Wunsch, 1996]. The solution rank – 8, out of 12 constraint equations – is selected to yield a dynamically acceptable solution in which perturbations to initial estimates of unknowns remain within a priori uncertainties. Overall, reference velocities are modified by mean (peak) perturbations of 2 (10) mm s$^{-1}$. The largest adjustments are introduced where observations are lacking, over Belgica Bank and north of Bear Island. Sea ice mean (peak) advection velocity adjustments are 2 (10) mm s$^{-1}$, equivalent to 0.5 mSv adjustments on average. The mean adjustment of surface FW flux is 20±88 mSv, with a peak of 150 mSv. The diagnosed diapycnal velocities have median values of 1-3×10$^{-7}$ m s$^{-1}$.

*c. Horizontal and diapycnal volume transports*

Fig.4 shows the spatial structure of the annual-mean Arctic boundary absolute velocity section, and full-depth and layer-specific volume transports. Volume transports (annual-mean ±1 sd) by gateway are summarized in Table 3. Davis Strait volume transport is –2.1±0.7 Sv, comprising –3.2±0.6 Sv outflow in the west (the Baffin Island Current) and 1.2±0.8 Sv inflow in the east (West Greenland Current). Most of the export is found in the SURF layer, –1.7±0.4 Sv. The Davis Strait volume transport seasonal cycle is similar to the estimate of Curry et al. [2014], −1.7±*0.5* Sv (±*uncertainty*), in same period of September 2005 – August 2006. In Fram Strait west of 6.5˚W, there is a cyclonic circulation of –0.4±0.5 Sv over Belgica Bank. There is –6.2±1.2 Sv export in the EGC between 6.5-2.0°W. In the WSC region east of 5.0°E, the volume transport is 7.4±1.0 Sv inflow, comprising 3.7±0.5 Sv AW, 1.4±0.4 Sv IW and 2.3±0.6 Sv DW (Fig. 4d). Overall Fram Strait volume transport is -1.1±1.2 Sv, comparable to the –2.0±2.7 Sv (±1 sd) of Schauer et al. [2008]. The circulation in the DW layers is almost closed (net transport is zero). The BSO volume transport is 2.3±1.2 Sv, comparable to 2.0 Sv mainly during 1997–2007 [Smedsrud et al., 2010], and is dominated by 1.3±1.0 Sv, mainly AW inflow in the middle region, and by 0.7±0.2 Sv inflow in the Norwegian Coastal Current. Bering Strait volume transport is 0.7±0.7 Sv, comparable to 0.8±*0.2* Sv (± *uncertainty*) during 1991–2004 [Woodgate et al., 2005]. Negative transports are seen in November (-0.5 Sv) and December (-0.4 Sv) 2005, but there are often wind-forced transport reversals between October and March [Woodgate et al. 2005].

Fig. 5a summarises the oceanic volume transport time series in the four major gateways. Their 12 monthly average and standard deviations are shown in Table 3. Net liquid ocean transport is -0.15±0.06 Sv. As prescribed in the inverse model, our volume budget is closed to within 1 mSv for each month. Net liquid ocean transports of –0.15±0.06 Sv combine with sea ice exports of –0.06±0.04 Sv to balance with the inverse model-derived surface FW input of 0.20±0.08 Sv (the 1 mSv difference in totals is due to rounding error).

Fig. 5b shows sea ice volume transport estimated across the four major gateways based on PIOMAS monthly output [Zhang and Rothrock, 2003]. Net (mean) sea ice export of -59±38 mSv is dominated by Fram Strait, at -51±34 mSv. Other contributions are small: 9±10 mSv (Davis Strait), and 1±12 mSv (Bering Strait). Sea ice export from the BSO is indistinguishable from zero. Our wintertime estimate (October 2005 – April 2006) of 85 mSv agrees with the 87 mSv estimate of Spreen et al. [2009] for the same period. Our annual mean estimate of 51±34 mSv is smaller than the 1990-1996 mean of 90 mSv [Vinje et al. 1998], but this difference may be due to the different observation periods, as late-winter sea ice thickness in Fram Strait has reduced from 4.3±0.4 m during 1990's to 2.0 m 2010 [Hansen et al. 2013, Renner et al. 2014]. For Davis Strait, our values agree with those of Curry et al. [2014]. For Bering Strait, Travers [2012] estimates sea ice export of 6±2 mSv for 2007-2008, which is consistent with our estimate.

The horizontal volume transports associated with each defined water masses (Fig. 4d) are balanced with inverse model derived diapycnal velocity and associated volume transports. Their shape is similar to that diagnosed in T2012, where 1.1±0.3 Sv upwells out of the AW layer upper surface, and 2.2±0.9 Sv downwells out of the AW layer lower surface. These diapycnal fluxes balance the horizontal convergence in the AW layer at rate of 3.7±1.1 Sv.

*d. Temperature and heat transports*

Key to the visualization of the oceanic temperature transports is the product $\theta'v'$, shown in Fig. 6, with associated full-depth and layer transports. The greatest constristions of $\theta'v'$ to the net heat transport appear in the upper ~200 m in western Davis Strait, over Belgica Bank and in the EGC, and the central BSO and Norwegian Coastal Current regions; and to greater depths, ~500 m, in the WSC. Integrated around the boundary, the liquid ocean component of the heat flux is 154±44 TW (annual mean ±1 sd). The dominant contribution arises from the AW (69±18 TW-eq), followed by SURF (47±21 TW-eq) and UAW (30±10 TW-eq) as shown in table 5.

Fig. 7a shows the time series of monthly oceanic heat transports and the contribution of each gateway to the total. There is a clear seasonal cycle in the heat transport, with highest values ~200 TW in September-January and lowest values ~120 TW in March-June. This variability mainly stems from the BSO and Fram Strait. Fig. 7b shows the net heat flux across the boundary as a combination of oceanic heat transport and sea ice latent plus sensible heat transports. The annual mean (±1 sd) heat transport is 175±48 TW. The amplitude and phase of the seasonal cycle are slightly modified by the addition of the sea ice component.

We next examine the contributions to seasonal variability in ocean heat transport by isolating the time-mean and time-varying contributions of temperature and velocity to total heat transport, following Lique et al. [2009]. For each of twelve months k (k=1,…,12), the oceanic heat transport ($F_{H,k}^o$) is decomposed into a time-mean component and three time-varying components, as follows.

$$F_{H,k}^o = \langle \Theta \rangle \langle V \rangle + V_k'\langle \Theta \rangle + \Theta_k'\langle V \rangle + \Theta_k'V_k' \qquad (3)$$

where $\Theta_k = \Theta_k(s,z) \equiv \theta_k'(s,z)$ is the boundary temperature anomaly field (for each month), $V_k = V_k(s,z) \equiv v_k'(s,z)$ is the boundary velocity anomaly field (for each month). Angle brackets indicate the pointwise time-mean (over the twelve months) of the relevant field, and primes now indicate the pointwise anomaly of each monthly field about the time-mean field. Products of pairs of terms on the RHS of (1) are taken to mean the integral of the product of the fields around the boundary with respect to *s* and *z*, following inversion to generate closed budgets. $\langle \Theta \rangle \langle V \rangle$ is the mean heat transport calculated as the integral of the product of the annual-mean velocity and temperature (anomaly) fields. $V_k'\langle \Theta \rangle$ is the component of heat transport resulting from monthly changes in advection of the annual-mean temperature field, and similarly $\Theta_k'\langle V \rangle$ is the component resulting from monthly changes of temperature advected by the annual-mean velocity field. Finally, $\Theta_k'V_k'$ is the correlation term of monthly potential temperature and velocity anomalies.

Fig. 7c shows all of the four components on the RHS of (3). $\langle \Theta \rangle \langle V \rangle$ = 148±1 TW, the absolute value is similar to the inverse model annual-mean heat transport of 154±44 TW. $V_k'\langle \Theta \rangle$ = 2±28 TW and $\Theta_k'\langle V \rangle$ = -2±25 TW make contributions of similar magnitude and phase to the seasonal variability in net heat transport, implying that variability in temperature and in velocity are of similar importance to oceanic heat transport variability. The correlation term $\Theta_k'V_k'$ = 7±4 TW has minor contribution on the seasonal cycle variability. We next examine these components in different water masses (Table 4). Inspection of temperature transport by layer shows that month-to-month variability in the AW layer and seasonal cycle in the SURF layer are the major sources of the temperature transport variability. The month-to-month temperature transport variability in the AW layer comes from the $V_k'\langle\Theta\rangle$ term – i.e. velocity variability. This velocity variability occurs mainly in Fram Strait and the BSO. The seasonal cycle in the SURF layer comes from the $\Theta_k'\langle V\rangle$ term. – i.e. temperature variability. This temperature variability arises from inflow regions around the boundary: the eastern part of Davis Strait, the WSC region in Fram Strait, BSO and Bering Strait, seen as higher temperature variability at 50 m depth in Fig. 2a.

We next re-calculate our temperature transports in each gateway to compare with previous estimates (annual mean ±1 sd). In the BSO, we find 64±33 TW-eq referenced to 0.0°C, consistent with the estimate of 73 TW-eq referenced to 0.0° C [Smedsrud et al., 2010]. Davis Strait temperature transport referenced to -0.1°C is 29±10 TW-eq, comparable to the annual mean estimate of 20±9 TW-eq (± *uncertainty*) during 2004-2005 [Curry et al., 2011]. Bering Strait temperature transport referenced to -1.9°C is 8±13 TW-eq. This is smaller than the most recent estimate of about 23 TW, which includes a correction of 3 TW-eq for the Alaskan Coastal Current, during August and October 2005, referenced to -1.9°C [Woodgate et al., 2010]. The WSC temperature transport referenced to -0.1°C is 58±10 TW-eq. This is larger than the recent annual mean temperature transport estimate of 28–44 TW-eq referenced to -0.1°C based on mooring observations during 1997–2000 [Schauer et al., 2004]. Lastly, Schauer et al. [2008] and Schauer and Beszczynska-Möller [2009] estimate decadal heat transports associated with AW inflow in Fram Strait since 1998 employing their "tube" method. They estimate heat transports of 40–50 TW during 2003–2007. We find heat transport associated with AW inflow in Fram Strait accounts for 26-32% of the 154±44 TW total oceanic heat transport.

*e. FW transports*

FW transports are visualised using $S'v'$, as for temperature transport, and the product is shown in Fig. 8 with associated full-depth and layer transports. The derived FW flux is dominated by the upper ~200 m, with large contributions in Davis Strait, Belgica Bank, the EGC and Bering Strait, and contributions from greater depths (~500 m) in the WSC. FW transport mirrors temperature transport to a large extent. The annual mean oceanic FW transport (±1 sd) is 155±65 mSv, with major contributions from the SURF (79±59 mSv-eq) and AW layers (57±15 mSv-eq) as shown in Table 5.

Fig. 9a shows the time series of net oceanic FW transport and the contribution of each gateway to the total. There is a clear seasonal cycle, with higher values ~250 mSv in November-January and lower values of ~80 mSv in August-September. Only the BSO contribution to the total is small; the transports in Davis Strait (109±13 mSv-eq) and Fram Strait (79±22 mSv-eq) are of similar magnitude; but the variability is clearly dominated by Bering Strait (-48±52 mSv-eq). Fig. 9b shows the boundary FW flux time series as the sum of oceanic and sea ice FW transports. The annual mean (±1 sd) is 204±85 mSv, with the ocean seasonal cycle modified by the smaller signal from the sea ice FW transport. Higher values of ~300 mSv are seen in November-January, and lower values of ~130 mSv in May-September.

We next pursue the same decomposition for monthly FW transports ($F^o_{FW,k}$) as for oceanic heat transports in (3) above:

$$F^o_{FW,k} = \langle\Sigma\rangle\langle V\rangle + V_k'\langle\Sigma\rangle + \Sigma_k'\langle V\rangle + \Sigma_k'V_k' \tag{4}$$

where now $\Sigma_k = \Sigma_k(s,z) \equiv S_k'(s,z)/\bar{S}_k$ is the boundary salinity anomaly field scaled by boundary mean salinity (for each month). Fig. 9c shows all of the four components on the RHS of (4). Again, the correlation term ($\Sigma_k' V_k'$) is small, but in contrast to heat transports, it is clear that oceanic FW transport variability is dominated by velocity variability ($V_k'\langle\Sigma\rangle$). The contribution of salinity variability ($\Sigma_k'\langle V\rangle$) is minor because the phase of the salinity variability is very similar in both inflow and outflow regions so that the contribution of salinity variability in each gateway cancels out. We next examine these components in different water masses (Table 5). Inspection of FW transport by layer shows that the FW transport variability is concentrated just in the SURF layer, driven by velocity variability. This velocity variability mainly stems from Bering Strait (Fig. 9a).

We next recalculate our results with different salinity reference values for comparison with previous studies (annual mean ±1 sd). Our Davis Strait FW transport is 117±14 mSv-eq, referenced to 34.8. Curry et al. [2014] estimates 97±*15* mSv-eq (± *uncertainty*) for the same period and the same reference value. In Fram Strait, de Steur et al. [2009] estimates the annual mean liquid FW outflow in the EGC region (6.5–0.0°W), referenced to 34.9, as 40.4±14.4 mSv-eq (±1 sd) based on 1998–2008 mooring observations. This value is similar to our liquid FW flux of 52±11 mSv-eq, for 6.5-2.0°W and the same reference salinity. The BSO FW flux referenced to 35.0 is -8±5 mSv-eq, compared to -17 mSv-eq by Smedsrud et al. [2010], who use mooring observations and a reference salinity of 35.0. The Bering Strait FW transport is 51±54 mSv-eq referenced to 34.8, which is smaller than the 80±*10* mSv-eq (± *uncertainty*) estimate of Woodgate and Aagaard [2005], based on 1991-2004 mooring observations. They include contributions from the Alaskan Coastal Current and salinity stratification, equivalent to 13 mSv-eq each. Although we include Alaskan Coastal Current and salinity stratification based on NEMO output, the sum of these contributions is small, 5±9 mSv-eq. This and the southward FW transport in November-December 2005 (Fig. 9a) may explain the discrepancies.

*f. Uncertainty of the total ocean heat and FW transports*

We have one important point of comparison for our new, mooring-based flux results: the calculations of T2012. They are of high spatial resolution and span the water column (nearly) from coast to coast and surface to sea-bed, but are quasi-synoptic. They are derived from hydrographic CTD measurements taken over 32 days from 9 August to 10 September 2005, and bottom mooring measurement within 2 months from 21 July to 27 September for the reference velocity (see T2012 Fig. 3). Therefore, we are aware that the overlap between our new results and those of T2012 is imperfect. See details of a comparison between first month mooring based estimate in September 2005 and T2012 for online supporting information B. Note that we also update the T2012 estimate including IBCAO bathymetry, updating the inverse model setting and correcting transport weighted seawater properties. Details are provided in online supporting information C.

Next, we examine the robustness of the total ocean heat and FW transports by investigating (first) the impact of the NEMO model output, where it is used to fill unobserved regions (the upper 50 m, and over the shallow shelves), and (second) the sparseness of the salinity measurements in Fram Strait. In short, we find that unobserved variability in the upper 50 m and over shelf regions affects the net oceanic heat transport by less than 7 TW (Fig. 10a). For the oceanic FW transport uncertainties, they stem mainly from the unobserved variability in the upper 50 m in western Davis Strait, Belgica Bank variability in Fram Strait, and sparse salinity measurements in the EGC region in Fram Strait (Fig. 10b-e). We estimate that their uncertainties are each ~30 mSv-eq. Improved observations in these regions would reduce uncertainty in FW transport estimates. See Appendix B for more detail.

*g. transport-weighted mean properties*

Water mass transformations can be interpreted as a result of air-sea heat and FW fluxes in outcrop regions at the sea surface and interior mixing [Walin 1982]. In the Arctic Ocean, Rudels et al. [2015] discuss the corresponding temperature and FW transports in Fram Strait based on water mass transformation in the Eurasian basin. Similarly, Pemberton et al. [2015] investigate the causes of water mass transformation in the Arctic Ocean based on cores (1˚) NEMO model output. Therefore, it is useful to translate our boundary heat and FW flux time series into water mass transformations in θ-S coordinates, to provide observation-based reference values for future studies.

As in T2012, monthly velocity fields are transformed from geographical coordinates (distance against pressure) into θ-S coordinates, gridded with $\Delta\theta = 0.2°C$ and $\Delta S = 0.05$. All transports within each θ-S grid box are summed and the net transports in the class are calculated. Based on the net volume transports ($m^3 s^{-1}$) per class, volume transport and seawater property change associated with the water mass transformation are calculated. These quantities are related to surface heat and FW fluxes (5, 6):

$$F_H^{surf} = \rho c_p (T_{in} - T_{out}^{oi}) V_{in} + F_H^{stor} \quad (5)$$

$$F_{vol}^{surf} = (S_{in} - S_{out}^{oi}) V_{in} / S_{out}^{oi} + F_{FW}^{stor} \quad (6)$$

where $V_{in}$ is inflow volume transport associated with the water mass transformation, $T_{in}$ and $S_{in}$ are transport-weighted inflow potential temperature and salinity, $T_{out}^{oi}$ and $S_{out}^{oi}$ are transport-weighted outflow temperature and salinity, including the sea ice contributions. These are alternative forms of (1) and (2); see Appendix C for derivation.

Fig. 11a shows time series of $V_{in}$, with annual mean (±1 sd) 11.3±1.2 Sv, and with a small seasonal cycle of ~1 Sv, higher during September-January and lower during February-May. Fig. 11b shows time series of $T_{in}$, $T_{out}^o$ and $T_{out}^{oi}$ with annual means (respectively; ±1 sd) 3.31±0.78°C, 0.03±0.21°C, -0.43±0.48°C, respectively. $T_{in}$ is highest in late summer and lowest in late winter. $T_{out}^o$ has much lower temporal variability than $T_{in}$; $T_{out}^o$ remains close to 0°C. Adding the sea ice contribution introduces a modest seasonal cycle, so that $T_{out}^{oi}$ has lower values than $T_{out}^o$ between autumn and late winter, by 0.3-0.7°C.

Fig. 11c shows time series of $S_{in}$, $S_{out}^o$, $S_{out}^{oi}$, with annual means (±1 sd) 34.57±0.18, 34.10±0.07, 33.95±0.09, respectively. $S_{in}$ has a clear seasonal cycle, with low values of ~34.4 during summer and higher values of ~34.7 during winter. $S_{out}^o$ has no clear seasonal cycle. Adding the sea ice contribution, this $S_{out}^o$ time variation is modulated during September to April, as seen in $S_{out}^{oi}$ time series.

Fig. 11d shows time series of potential density in inflow ($D_{in}$), outflow without ice ($D_{out}^o$) and with sea ice ($D_{out}^{oi}$), which are calculated based on $T_{in}$, $T_{out}^o$, $T_{out}^{oi}$, $S_{in}$, $S_{out}^o$, $S_{out}^{oi}$. Annual means (±1 sd) of $D_{in}$, $D_{out}^o$, $D_{out}^{oi}$ are 27.51±0.18, 27.37±0.06, 27.28±0.06 kg m$^{-3}$, respectively. $D_{in}$ has a seasonal cycle, with high values of ~27.7 during winter, and lower values of ~27.4 during summer, reflecting the seasonal cycles of $T_{in}$ and $S_{in}$. Density variability is dominated by salinity variability at low temperatures when the salinity range is (relatively) large.

Fig. 11e shows the monthly boundary heat fluxes (section 3d, Fig. 7b) with the diagnosed time series based on equation (4). These two time series are very similar (as they should be). The boundary heat flux is higher (~220 TW) during September – January because of the larger temperature difference between $T_{in}$ and $T_{out}^{oi}$ (~4°C) and stronger volume inflow ($V_{in}$~12 Sv), and it is lower (~130 TW) during March – June because of the smaller temperature difference between $T_{in}$ and $T_{out}^{oi}$ of (~3°C) and weaker volume inflow ($V_{in}$~10 Sv).

Fig. 11f shows the time series of net boundary FW flux (section 3e, Fig. 10b) with the diagnosed time series from equation (5). The net FW flux is higher (~300 mSv) during early winter as a result of the larger salinity difference between $S_{in}$ and $S_{out}^{oi}$ (~0.9) and larger seawater volume transports. It is lower (~150 mSv) during summer as a result of the smaller salinity difference between $S_{in}$ and $S_{out}^{oi}$ (~0.4) and smaller seawater volume transport.

We next view the net θ-S transformations of monthly inflows into outflows on θ-S space (Fig. 12). Considering first the all-liquid inflow, we note that the seasonal cycles of liquid seawater temperature and salinity are roughly in quadrature (i.e. ~3 months out of phase). Salinity extrema occur in summer (winter), which we associate with maximum (minimum) insolation causing maximum rates of melting (freezing). Temperature extrema occur in autumn (spring), when heat has ceased (begun) to be input to the ice and ocean. The effect is that the mean properties rotate (roughly) clockwise in θ-S phase space through the year. In contrast, the liquid outflow is more compressed in θ-S space, with the property ranges reduced compared with inflow values. It is notable that the highest seawater salinities are seen in late winter, as would be expected from the net impact of brine rejection by the end of the freezing period. If sea ice is included in the outflow properties (via temperature scaling based on specific and latent heat capacities), then the mean outflow temperatures and salinities are both reduced, and the range of temperatures is increased, while the range of salinities is not. This latter point likely reflects the assumption of constant sea ice salinity.

Based on the differences between the annual means of in- and outflows, the annual mean net effect of the Arctic is to freshen and cool the inflows by 0.62±0.23 in salinity (including sea ice) and 3.74±0.76°C, and there is a net input of buoyancy shown by the decrease in mean density (including sea ice) of 0.23±0.20 kg m$^{-3}$.

## 4. Discussion

In this section, we discuss the meaning and implications of the results presented in section 3. We first consider the meaning of our results in the light of the absence of direct estimates of interior storage fluxes. We next compare our results with atmospheric re-analysis and the NEMO model.

*a. Fluxes and storage*

We have generated time series of twelve monthly values, spanning a year, of heat and FW fluxes around the defined Arctic ice and ocean boundary. In this section, we discuss the meaning of these time series in terms of surface fluxes, in the absence of complementary estimates of heat and FW storage.

Storage fluxes of heat and FW are undoubtedly important in the Arctic Ocean, both on seasonal and longer timescales, but they cannot be yet be estimated directly by means of in situ census. The technologies available to make year-round property measurements – moorings and ice-tethered profiling instruments – are not yet sufficiently spatially dense to resolve changes at the monthly to seasonal timescale. Various authors have described long-term, of order decadal, changes in FW storage, and Rabe et al. [2014] argue that annual resolution is possible. We also note that there are clear signs of the emergence of the seasonal cycle in FW storage from the "background" in recent years [Rabe et al. 2014, supplementary material, Fig. 3]. Nevertheless, we can estimate the seasonal cycle of FW storage from measurements using remote-sensed altimetry and gravimetry in combination to calculate mass and steric contributions to total sea surface height changes, and then assuming, as a consequence of the low-temperature environment, that the steric changes are due to salinity and hence FW storage changes. It is reasonable to approximate the seasonal cycle of FW storage, therefore, as the sum of two components: a repeating seasonal cycle of zero mean, and a long-term trend [Armitage et al. 2016, their Fig. 5]. As a consequence, the annual average of the ice and ocean boundary FW flux accurately represents the annual average of the surface flux of FW, when the long-term trend is included as a relatively small contribution to its uncertainty.

We can assert, by analogy, that the same holds true for heat storage within the control volume: the annual average of the ice and ocean boundary heat flux accurately represents the annual average of the surface heat flux. However, it is more difficult to justify this assertion with reference to measurements, because the focus of much recent work on storage has been on FW and not on heat. Some studies have taken a long-term view of Arctic temperature changes, such as Steele and Boyd [1998], Korhonen et al. [2013] and Polyakov et al. [2012 & 2017]. We make a scale calculation for the long-term heat storage flux as follows. Assume that 10% of the Arctic area is affected ($10^{12}$ m$^2$), and that a depth range of 500 m is warmed by 0.5 °C over 10 years; with density 1000 kg m$^{-3}$ and heat capacity 4000 J kg$^{-1}$ °C$^{-1}$, the resulting heat flux is 3 TW, which is negligible. Mayer et al. [2016] also estimate Arctic Ocean has been accumulated heat at rate of ~1 Wm$^{-2}$ (~10 TW) over last decades based on historical hydrographic observations and data assimilation model outputs. However, the seasonal cycle in surface heat flux is certainly not negligible. Can we reach a conclusion about the meaning of the ice and ocean boundary view of heat fluxes?

We know that the long-term means of surface and boundary heat fluxes must be the same, because the surface heat fluxes cool (freeze) and warm (melt) the seawater (sea ice): warm waters enter the Arctic Ocean, and cold waters and sea ice leave. Now Bacon et al. [2015] illustrate, using NEMO model output, the large disparity in the amplitudes of the surface and boundary seasonal cycles, for the same control volume: the surface heat flux amplitude is ~500 TW, the boundary amplitude ~50 TW. However, we note that almost all waters entering the Arctic take a very long time to reach an exit: from several years to decades, and even to centuries. For the surface and halocline waters, the transient times are of between 2-16 years [Schlosser et al. 1999; Ekwurzel et al. 2001]. For the Atlantic Water the time to travel along the Arctic margins until the Atlantic Water exits the Arctic Ocean through the western Fram Strait is $O(20)$ years [Mauldin et al. 2010; Karcher et al. 2011, 2012] and for the Pacific inflow the transient times are $O(10)$ years [e.g., Aksenov et al., 2016]. Renewal times for the deep Canada Basin have been estimated at centuries [Timmermans and Garrett 2006]. The exception is the Fram Strait recirculation, where waters entering the volume on western side of the West Spitzbergen Current may only spend weeks– months inside the control volume before leaving again as part of the East Greenland Current. Therefore the seasonal cycle of surface heat (and FW) flux is smoothed out during the long residence times within the control volume, through the application of many (maybe tens or even hundreds of) seasonal cycles.

These are our conclusions as to the meaning of the ice and ocean boundary fluxes. First, the annual mean boundary fluxes accurately represent the relevant annual mean surface fluxes, even in the absence of storage measurements. Second, we use the phrase "accurately represent" rather than "are equal to" because the boundary fluxes are the result of a complex convolution of the trajectories of individual water parcels with the action of surface fluxes upon them over many years. Third, the individual monthly boundary fluxes do not represent surface fluxes, in the absence of storage measurements.

*b. Comparison with atmospheric re-analysis and NEMO model*

The annual mean (±1 sd) boundary heat flux estimate is 175±48 TW, equivalent to 15.5±4.2 $Wm^{-2}$. According to Mayer et al. [2016], that the Arctic Ocean heat content has been increasing at rate of ~1 $Wm^{-1}$ during 2000-2015. Assuming that month-to-month heat storage variability becomes negligible by averaging over a year, sum of these two numbers can be compared to long-term average surface net heat fluxes. Porter et al. [2010] and Cullather and Bosilovich [2012] update Serreze et al. [2007b]'s atmospheric heat budget estimate with atmospheric re-analysis output and satellite-based estimates. Note that their estimates refer to the polar cap north of 70°N, and they include most of the Nordic Seas, where large surface heat flux happens. Their long-term mean surface heat fluxes range from 5 $Wm^{-2}$ for the NCEP / NCAR Reanalysis [NRA; Kalnay et al., 1996], 11 $Wm^{-2}$ for the ECMWF Reanalysis [ERA-40; Uppala et al., 2006], 14 $Wm^{-2}$ for the Contemporary re-analysis from the Climate Forecast Center [CFSR; Saha et al. 2010], to 19 $Wm^{-2}$ for the Modern-Era Retrospective Analysis for Research and Applications [MERRA; Cullather and Bosilovich, 2011]. CFSR and MERRA are closest to our estimate.

We do similar assessment on surface FW flux. The annual mean (±1 sd) boundary FW flux of 204±85 mSv (6,430±2,680 $km^3 yr^{-1}$). The long-term trend of FW content increase within the side boundary is estimated as ~10 mSv (~315 $km^3 yr^{-1}$), considering recent liquid FW content increase during 1992-2012 [Rabe et al., 2014] and decline of sea ice over the last decades [Lindsay and Schweiger et al., 2015]. The sum of them agrees well with the surface FW flux estimate of Haine et al. [2015] of 6,770 km$^3$ yr$^{-1}$ for the period of 2000-2010, which is the sum of river runoff of 4,200±*420* km$^3$ yr$^{-1}$ (± *uncertainty*), excess of precipitation over evaporation of 2,200±*220* km$^3$ yr$^{-1}$ (± *uncertainty*), and FW input from Greenland into the Baffin Bay of 370±*25* km$^3$ yr$^{-1}$ (± *uncertainty*). The agreement is very good, even though about a third of the Arctic FW runoff volume is ungauged [Shiklomanov and Lammers, 2009]. Precipitation estimates differ in the seven most recent atmospheric re-analysis models [Lindsay et al., 2014], and this is one of key processes in the atmosphere that requires better quantification and understanding to enable better predictions of the Arctic freshwater system [Vihma et al., 2016].

Pemberton et al. [2015] provide another useful point of comparison by diagnosing annual-mean water mass transformations in a low-resolution (1º) version of the NEMO model run. They use a similar domain to ours, with gateways at Bering and Fram Straits and the BSO; however, instead of Davis Strait, they close the boundary at the model's Barrow and Nares Straits, in the Canadian Arctic Archipelago, thereby omitting Baffin Bay, for which we must make some allowance. Their transport-weighted inflow and outflow temperatures, salinities and densities (compared with ours) are $T_{in}$ = 2.9 (3.3) ºC, $T^o_{out}$ = -0.7 (0.0) ºC, $S_{in}$ = 34.3 (34.6), $S^o_{out}$ = 33.6 (34.1), and $D_{in}$ = 27.34 (27.51), $D^o_{out}$ = 27.01 (27.37). Thus, Pemberton et al. [2015] cool and freshen the inflow by 3.6 ºC and 0.7, compared with our 3.3 ºC and 0.5 (respectively), resulting in a density change (towards lower density) of 0.33, more than double our estimate of 0.14, which results mainly from their stronger freshening. Over-freshening of outflows could have important down-stream consequences for the MOC, for example. The omission by Pemberton et al. [2015] of Baffin Bay makes little difference to these property changes because Baffin Bay is only ~5% of our domain's total surface area, while its surface heat [Aksenov et al. 2010a] and FW fluxes [Haine et al. 2015] are similar to Arctic net values.

## 5. Summary

We have presented, for the first time, the seasonal variability of Arctic ice and ocean boundary heat and FW fluxes, expressed as 12 monthly mean fluxes from September 2005 to August 2006 (Fig. 5, 7, 9 and Table 3). Their annual means (±1 sd) are 175±48 TW and 204±85 mSv, and they include sea ice contributions of 22±15 TW and 48±32 mSv respectively. They compare reasonably well with models, re-analysis and data compendia. Furthermore, these annual mean boundary fluxes are equivalent to annual mean surface fluxes, and they are the first (almost) entirely measurement-based estimates of these quantities.

The boundary heat flux variability derives mainly from velocity variability in the AW layer and temperature variability in the surface layer (Table 4). We represent unobserved variability in the upper 50 m and over shelf regions with model output, but this affects the net oceanic heat transport by less than 7 TW (Fig. 10a). We find that the FW flux variability is dominated by Bering Strait FW transport variability, which in turn is dominated by its velocity variability (Table 5 and Fig. 9a). The oceanic FW transport uncertainties stem mainly from the unobserved variability in the upper 50 m in western Davis Strait, Belgica Bank variability in Fram Strait, and sparse salinity measurements in the EGC region in Fram Strait (Fig. 10b-e). We estimate that their uncertainties are each ~30 mSv-eq. Improved observations in these regions would reduce uncertainty in FW transport estimates.

The boundary flux estimates are converted into associated water mass property changes, as transport-weighted temperature and salinity variability in the inflow and outflow (Fig. 11, 12). Inflow temperature and salinity have clear seasonal cycles but different phase, while outflow temperature varies little, and outflow salinity has no clear seasonal cycle. The annual net effect of the Arctic is to freshen and cool the inflows by 0.62±0.23 in salinity and 3.74±0.76°C in temperature, with a resulting net input of buoyancy shown by the decrease in mean density of 0.23±0.20 kg m$^{-3}$.

*Acknowledgments:* This study was funded by the UK Natural Environment Research Council as a contribution to the Arctic project TEA-COSI (NE/I028947/1, PI Bacon). The Davis Strait mooring data were collected by UW, Seattle, USA (PI Lee) with funding from NSF grant (ARC-1022472). The data and objectively mapped sections are available via http://iop.apl.washington.edu/data.html. The Bering Strait mooring data were collected by UW, Seattle, the USA and the UAF, Fairbanks, the USA (PIs Woodgate and Weingartner) with funding from NSF grant (NFS-0856786), and NOAA- RUSALCA program. The data are freely available via http://psc.apl.washington.edu/BeringStrait.html. The BSO mooring data were collected by IMR, Bergen, Norway (PI Ingvaldsen). The Fram Strait mooring data were collected by AWI, Bremerhaven, Germany, and NPI, Tromsø, Norway (PIs Schauer and Hansen). The central and eastern Fram Strait mooring data are available via https://doi.pangaea.de/10.1594/PANGAEA.845938. The BSO and western Fram Strait mooring data is available on request. The volume, heat and FW transport time series presented in this manuscript can be accessed via https://doi.pangaea.de/10.1594/PANGAEA.870607.

# APPENDIX A

Setting of the box inverse model

There are 1,287 unknowns to be determined by the inverse model. The high horizontal resolution of 3 km requires 639 values of horizontal velocity offsets to be determined, and the same number of values of sea ice velocity. There are 4 layer interfaces where diapycnal transports of volume and salinity velocity are to be determined. The surface FW input is one further unknown.

We define 5 layers bounded by isopycnal surfaces (Table 2): Surface Water (SURF) layer, Upper Atlantic Water (UAW) layer, Atlantic Water (AW) layer, Intermediate Water (IW) layer, Deep Water (DW) layers. Water mass definitions are the same as in T2012, with one change: we now define Surface Water as the combination of T2012's two uppermost layers (Surface Water and Subsurface Water). The following flux constraints are applied: full depth conservation of volume and salinity transports (1 constraint each), and volume and salinity transports for each layer (5 constraints for each), a total of 12 constraints. We prescribe salinity conservation instead of salinity anomaly conservation because the model solution can be distorted by large FW transport imbalances appearing in salinity anomaly constraints, whereas salinity constraints generate stable solutions.

The a priori reference velocity uncertainty is estimated as the 3-month standard deviation of moored velocity measurements (0.01-0.05 m s$^{-1}$). Uncertainties are linearly interpolated onto each station pair accordingly. Following T2012, larger a priori uncertainties are associated with Belgica

Bank (0.06 m s$^{-1}$) and the northern BSO, where moored velocity measurements are lacking. Bering Strait uncertainties are set to 10% of its 3-month standard deviation (~0.02 m s$^{-1}$), to take into account (a) its larger short-term variability, and (b) its longer-term variability is well observed [Woodgate et al. 2005]. The a priori uncertainty in the diapycnal velocities is set to $1 \times 10^{-5}$ m s$^{-1}$, near the upper end of the range of vertical velocities inferred from observed ocean mixing rates. The a priori uncertainty in the sea ice advection velocity is set to 10% magnitude of its initial estimate, unless sea ice thickness is less than 0.3 m, when the a priori uncertainty is set to zero to avoid large column weightings. The a priori uncertainty of surface FW input is set to 100% magnitude of its initial estimate.

All unknowns are initialized as follows. Horizontal ocean velocity offsets are initialized using moored velocity measurements or NEMO modeled velocities, as described above. Sea ice initial velocities are derived from PIOMAS monthly output by projection onto the inverse model grid. Diapycnal velocities are set to zero. The initial surface FW input is set to 180 mSv throughout the year. This is because we aim to diagnose this term from boundary measurements using the box inverse model. An initial sensitivity test shows that we can infer this term from boundary measurements up to 800 mSv.

## APPENDIX B

Uncertainty of the total ocean heat and FW transports

We examine the robustness of the total ocean heat and FW transports by investigating (first) the impact of the NEMO model output, where it is used to fill unobserved regions (the upper 50 m, and over the shallow shelves), and (second) the sparseness of the salinity measurements in Fram Strait.

To quantify the impact of NEMO, four different sets of monthly temperature, salinity and velocity fields are prepared. The first is called "Mooringonly", and uses no NEMO output, so we assume no stratification above the shallowest instruments, which means the upper 100m in central Davis Strait and the upper 50m in Fram and Bering Strait. For the shelves (Belgica Bank in Fram Strait and north of Bear Island in BSO), we put zero velocities. For temperature and salinity over the shelves, we apply uniform temperature and salinity profiles over the region. The second is the "Best estimate" dataset, as prepared in section 2b. The third is called "Inc50mNoShelf", and it includes the upper ocean variability in Davis, Fram and Bering Straits, but it does not have shelf region variability. The fourth is the "No50mIncShelf" estimate; it has no upper-ocean temperature, salinity or velocity variability (outside the shelf regions), but it includes the shelf region variability. Inversions with volume and salinity constraints for each layer and for full depth (as described in appendix A) are performed to obtain volume- and salinity-conserved velocity fields for all data sets.

Oceanic heat transports (Fig. 10a) are relatively insensitive to our permutations, with a range of ca. ±5 *TW* per month. For ocean FW transports (Fig. 10b), the shape of the month-to-month variability is similar across the various permutations, which means that the moored observations alone capture the majority of the FW transport seasonal cycle. However, there is a mean offset between the "Best estimate" and "MooringOnly" runs of 6±13 mSv. This is a result of competition between the impacts of the upper 50 m, where the "Inc50mNoshelf" runs add 27 mSv on average to the "Mooringonly" estimates, and of the shelf waters, where the "No50mIncShelf" runs subtract 32 mSv. To investigate the source of these offsets, we inspect the cumulative FW transport anomaly (with respect to the "Mooringonly" runs) around the boundary (Fig. 10c). The positive contribution to the FW transport from the upper 50 m almost entirely occurs in western Davis Strait (33±7 mSv-eq). Inclusion of Belgica Bank via the "No50mIncShelf" run has a substantial impact: the recirculation over the bank has an amplitude of ~60 mSv-eq, but it leaves a residual contribution to the FW transport of -27±6 mSv-eq. Improved observations in these regions would lead to better estimates of FW transports.

Next, we examine the impact on FW transports of the sparseness of the salinity measurements in Fram Strait by conducting a suite of sensitivity tests. Four different monthly sets of Fram Strait temperature and salinity fields are prepared as follows. First, an annual mean salinity section is generated from the monthly "Bestestimate" fields; this is called "MooringTSfix". Second, we use the monthly "Bestestimate" dataset itself. Third, we use the high-resolution, full-depth, hydrographic temperature and salinity fields from T2012, fixed throughout the year; this is called "T2012TSfix". Fourth, we generate high-resolution, time-varying temperature and salinity fields by temporal extrapolation through the year, starting with the T2012 temperature and salinity fields, and using the NEMO salinity temporal variability; this is the "T2012TSvar" estimate. Fig. 10d shows the four resulting sets of net oceanic FW transports across the pan-Arctic boundary. All the time series show similar seasonal cycles in terms of phase and magnitude. The major difference arises between pairs of runs. For one pair, there is little difference between use of the annual mean salinity section versus monthly salinity fields ("MooringTSfix" and "Bestestimate"). For the other pair, there is little difference between the use of the fixed T2012 salinity field and the same field with added NEMO temporal variability. However, there is a difference between the pairs of ~30 mSv, with the fields using T2012 salinity generating lower FW transport. Fig. 10e shows that the discrepancy comes from EGC region in Fram Strait. This is primarily due to sparse mooring

-35-

measurements, which fail to capture the halocline layer between 50-150 m water depth in the region. Comparison of the September 2005 mooring salinity section with CTD section during 26 August to 9 September 2005 highlights up to 1.0 salinity discrepancy in depths of 50-200 m in the region, where the mooring salinities appear too fresh. The mooring array fails to capture the salinity 34.5 halocline layer at depths of 50-150 m, so that fresher water is present as a result of the linear interpolation. More salinity observations in the EGC region (50-500 m) would lead to better estimates of FW transport. We suggest that ocean FW transport uncertainty associated with sparse salinity measurements in Fram Strait is ca. *±30 mSv*.

## APPENDIX C

Deviation of volume transport weighted temperature and salinity both in inflow and outflow associated with the water mass transformation

Monthly velocity fields on geographical coordinate (distance against pressure) are transformed into θ-S plane, gridded with dθ = 0.2°C and dS = 0.05. All transports within the same θ-S grid box are summed and the net transports in the class are calculated. Calculated net volume transports ($m^3 s^{-1}$) binned in θ-S class are stored in $\Pi(\theta, S)$.

Using the $\Pi(\theta, S)$, volume transport in inflow associated with the water mass transformation ($V_{in}$) is calculated.

$$V_{in} = \iint \delta_{in}(\theta, S) \Pi(\theta, S) dS d\theta \qquad (S1)$$

Where $\delta_{in}(\theta, S)$ is a box car function. It is 1 when $\Pi(\theta, S)$ is positive, and it is 0 when $\Pi(\theta, S)$ is negative. dS is salinity element and dθ is potential temperature element on the θ-S plane.

Volume transport weighted potential temperature in inflow ($T_{in}$) and volume transport weighted salinity in inflow ($S_{in}$) are calculated.

$$T_{in} = \iint \delta_{in}(\theta, S)\Pi(\theta, S)\theta dS d\theta / V_{in} \qquad (S2)$$

$$S_{in} = \iint \delta_{in}(\theta, S)\Pi(\theta, S) S dS d\theta / V_{in} \qquad (S3)$$

Oceanic volume transport in outflow associated with the water mass transformation without sea ice ($V_{out}^o$) and with sea ice ($V_{out}^{oi}$) are calculated.

$$V_{out}^o = \iint \delta_{out}(\theta, S)\Pi(\theta, S) dS d\theta \qquad (S4a)$$

$$V_{out}^{oi} = V_{out}^o + V_{out}^i \qquad (S4b)$$

Where $\delta_{out}(\theta, S)$ is a box car function. It is 1 when $\Pi(\theta, S)$ is negative, and it is 0 when $\Pi(\theta, S)$ is positive. $V_{out}^i$ is net sea ice export in each month, obtained from the PIOMAS output.

Volume transport weighted potential temperature in outflow without sea ice ($T_{out}^o$) and with sea ice ($T_{out}^{oi}$) are calculated.

$$T_{out}^o = \iint \delta_{out}(\theta, S)\Pi(\theta, S)\theta dS d\theta / V_{out}^o \qquad (S5a)$$

$$T_{out}^{oi} = \left\{\iint \delta_{out}(\theta, S)\Pi(\theta, S)\theta dS d\theta + \frac{\rho_i c_p^i}{\rho_o c_p^o}\left(-\frac{c_f}{c_p^i} + \theta^i\right) V_{out}^i\right\} / (V_{out}^o + V_{out}^i) \qquad (S5b)$$

Where $\theta^i$ is averaged sea ice temperature in each month, obtained from the PIOMAS output. $\rho_o$ is density of sea water (1,027 kg m$^{-3}$), $c_p^o$ is specific heat capacity of sea water (3.987*10$^3$ J kg$^{-1}$ K$^{-1}$), $\rho_i$ is density of sea ice (930 kg m$^{-3}$), $c_p^i$ is specific heat capacity of sea ice from Ono [1967] as a function of sea ice temperature and salinity, $c_f$ is heat of fusion (3.347*10$^5$ J kg$^{-1}$).

In a similar manner, volume transport weighted salinity in outflow without sea ice ($S_{out}^o$) and with sea ice ($S_{out}^{oi}$) are calculated.

$$S_{out}^o = \iint \delta_{out}(\theta, S)\Pi(\theta, S) S dS d\theta / V_{out}^o \qquad (S6a)$$

$$S_{out}^{oi} = \left\{\iint \delta_{out}(\theta, S)\Pi(\theta, S) S dS d\theta + S_i V_{out}^i\right\} / (V_{out}^o + V_{out}^i) \qquad (S6b)$$

Where $S_i$ is sea ice salinity in each month, which is set as 6 throughout the year.

Volume, heat and salt conservation equations can be described using above defined quantities (equation S1-S6).

$$\dot{V} = [V_{in} - V_{out}^{oi}] + F_{vol}^{surf} \qquad (S7)$$

$$\frac{\partial}{\partial t}\left(\sum H^{oi}\right) = \left[\rho_o c_p^o V_{in} T_{in} - \rho_o c_p^o V_{out}^{oi} T_{out}^{oi}\right] + \rho_F c_p^F F_{vol}^{surf} T_F - F_H^{surf} \qquad (S8)$$

$$\frac{\partial}{\partial t}\left(\sum B^{oi}\right) = V_{in} S_{in} - V_{out}^{oi} S_{out}^{oi} \qquad (S9)$$

Where $\dot{V}$ is change of volume in sea water and sea ice within the boundary, $F_{vol}^{surf}$ is surface FW input. $\sum H^{oi}$ is the heat stored in the ice-ocean system within the boundary, $\rho_F c_p^F F_{vol}^{surf} T_F$ is heat flux associated with the surface FW input, $F_H^{surf}$ is air-sea surface heat flux. $\sum B^{oi}$ is the salt stored in the ice-ocean system within the boundary.

Based on equation S7-S9 and neglecting contribution of heat flux associated with the surface FW input and $\dot{V}$, $F_{vol}^{surf}$ and $F_H^{surf}$ can be described as follows.

$$F_H^{surf} = \rho_o c_p^o (T_{in} - T_{out}^{oi}) V_{in} - \frac{\partial}{\partial t}\left(\sum H^{oi}\right) \qquad (S10)$$

$$F_{vol}^{surf} = (S_{in} - S_{out}^{oi}) V_{in}/S_{out}^{oi} + \dot{V} - \frac{1}{S_{out}^{oi}}\left[\frac{\partial}{\partial t}\left(\sum B^{oi}\right)\right] \qquad (S11)$$

# References

Aksenov, Y., S. Bacon, A. C. Coward, and N. P. Holliday, 2010a: Polar outflow from the Arctic Ocean: A high resolution model study. *J. Mar. Sys.*, **83,** 14-37, https://doi.org/10.1016/j.jmarsys.2010.06.007.

Aksenov, Y., S. Bacon, A. C. Coward, and A. J. G. Nurser, 2010b: The North Atlantic inflow to the Arctic Ocean: High-resolution model study. *J. Mar. Sys.*, **79,** 1-22, https://doi.org/10.1016/j.jmarsys.2009.05.003.

Aksenov, Y., and Coauthors, 2011: The Arctic Circumpolar Boundary Current. *J. Geophys. Res.*, **116**, https://doi.org/10.1029/2010jc006637.

Aksenov, Y., and Coauthors, 2016: Arctic pathways of Pacific Water: Arctic Ocean Model Intercomparison experiments. *J. Geophys. Res.*, **121,** 27-59, https://doi.org/10.1002/2015JC011299.

Armitage, T. W. K., S. Bacon, A. L. Ridout, A. A. Petty, S. Wolbach, and M. Tsamados, 2017: Arctic Ocean surface geostrophic circulation 2003-2014. *Cryosphere*, **11,** 1767-1780, https://doi.org/10.5194/tc-11-1767-2017.

Armitage, T. W. K., S. Bacon, A. L. Ridout, S. F. Thomas, Y. Aksenov, and D. J. Wingham, 2016: Arctic sea surface height variability and change from satellite radar altimetry and GRACE, 2003-2014. *J. Geophys. Res.*, **121,** 4303-4322, https://doi.org/10.1002/2015jc011579.

Bacon, S., Y. Aksenov, S. Fawcett, and G. Madec, 2015: Arctic mass, freshwater and heat fluxes: methods and modelled seasonal variability. *Phil. Trans. Roy. Soc.*, **373**, https://doi.org/10.1098/rsta.2014.0169.

Beszczynska-Moller, A., E. Fahrbach, U. Schauer, and E. Hansen, 2012: Variability in Atlantic water temperature and transport at the entrance to the Arctic Ocean, 1997-2010. *ICES J. Mar. Sci.*, **69,** 852-863, https://doi.org/10.1093/icesjms/fss056.

Carmack, E., and Coauthors, 2015: Toward Quantifying the Increasing Role of Oceanic Heat in Sea Ice Loss in the New Arctic. *Bull. Amer. Meteor. Soc.*, **96,** 2079-2105, https://doi.org/10.1175/Bams-D-13-00177.1.

Carmack, E. C., and Coauthors, 2016: Freshwater and its role in the Arctic Marine System: Sources, disposition, storage, export, and physical and biogeochemical consequences in the Arctic and global oceans. *J. Geophys. Res.*, **121,** 675-717, https://doi.org/10.1002/2015jg003140.

Cowtan, K., and R. G. Way, 2014: Coverage bias in the HadCRUT4 temperature series and its impact on recent temperature trends. *Quart. J. Roy. Meteor. Soc.*, **140,** 1935-1944, https://doi.org/10.1002/qj.2297.

Cullather, R. I., and M. G. Bosilovich, 2012: The Energy Budget of the Polar Atmosphere in MERRA. *J. Climate*, **25,** 5-24, https://doi.org/10.1175/2011jcli4138.1.

Curry, B., C. M. Lee, and B. Petrie, 2011: Volume, Freshwater, and Heat Fluxes through Davis Strait, 2004-05. *J. Phys. Oceanogr.*, **41,** 429-436, https://doi.org/10.1175/2010jpo4536.1.

Curry, B., C. M. Lee, B. Petrie, R. E. Moritz, and R. Kwok, 2014: Multiyear Volume, Liquid Freshwater, and Sea Ice Transports through Davis Strait, 2004–10. *J. Phys. Oceanogr.*, **44,** 1244-1266, https://doi.org/10.1175/JPO-D-13.

de Steur, L., E. Hansen, R. Gerdes, M. Karcher, E. Fahrbach, and J. Holfort, 2009: Freshwater fluxes in the East Greenland Current: A decade of observations. *Geophys. Res. Lett.*, **36***,* https://doi.org/10.1029/2009gl041278.

de Steur, L., E. Hansen, C. Mauritzen, A. Beszczynska-Möller, and E. Fahrbach, 2014: Impact of recirculation on the East Greenland Current in Fram Strait: Results from moored current meter measurements between 1997 and 2009. *Deep Sea Res.*, **92,** 26-40, https://doi.org/10.1016/j.dsr.2014.05.018.

Dickson, R., B. Rudels, S. Dye, M. Karcher, J. Meincke, and I. Yashayaev, 2007: Current estimates of freshwater flux through Arctic and subarctic seas. *Prog. Oceanogr.*, **73,** 210-230, https://doi.org/10.1016/j.pocean.2006.12.003.

Ekwurzel, B., P. Schlosser, R. A. Mortlock, R. G. Fairbanks, and J. H. Swift, 2001: River runoff, sea ice meltwater, and Pacific water distribution and mean residence times in the Arctic Ocean. *J. Geophys. Res.*, **106,** 9075-9092, https://doi.org/10.1029/1999jc000024.

Fahrbach, E., and P. Lemke, 2005: ARK-XX/2 cruise report, 57-142 pp, Place, Published, http://epic.awi.de/28681/.

Giles, K. A., S. W. Laxon, A. L. Ridout, D. J. Wingham, and S. Bacon, 2012: Western Arctic Ocean freshwater storage increased by wind-driven spin-up of the Beaufort Gyre. *Nature Geo.*, **5,** 194-197, https://doi.org/10.1038/Ngeo1379.

Haine, T. W. N., and Coauthors, 2015: Arctic freshwater export: Status, mechanisms, and prospects. *Global and Planetary Change*, **125,** 13-35, https://doi.org/10.1016/j.gloplacha.2014.11.013.

Hansen, E., and Coauthors, 2013: Thinning of Arctic sea ice observed in Fram Strait: 1990-2011. *J. Geophys. Res.*, **118,** 5202-5221, https://doi.org/10.1002/jgrc.20393.

Ingvaldsen, R. B., L. Asplin, and H. Loeng, 2004: The seasonal cycle in the Atlantic transport to the Barents Sea during the years 1997-2001. *Cont. Shelf Res.*, **24,** 1015-1032, https://doi.org/10.1016/j.csr.2004.02.011.

Jakobsson, M., 2002: Hypsometry and volume of the Arctic Ocean and its constituent seas. *Geochemistry Geophysics Geosystems*, **3,** 1-18., https://doi.org/10.1029/2001gc000302.

Karcher, M., A. Beszczynska-Moller, F. Kauker, R. Gerdes, S. Heyen, B. Rudels, and U. Schauer, 2011: Arctic Ocean warming and its consequences for the Denmark Strait overflow. *J. Geophys. Res.*, **116**, https://doi.org/10.1029/2010jc006265.

Karcher, M., J. N. Smith, F. Kauker, R. Gerdes, and W. M. Smethie, 2012: Recent changes in Arctic Ocean circulation revealed by iodine-129 observations and modeling. *J. Geophys. Res.*, **117**, https://doi.org/10.1029/2011jc007513.

Korhonen, M., B. Rudels, M. Marnela, A. Wisotzki, and J. Zhao, 2013: Time and space variability of freshwater content, heat content and seasonal ice melt in the Arctic Ocean from 1991 to 2011. *Ocean Science*, **9,** 1015-1055, https://doi.org/10.5194/os-9-1015-2013.

Kovacs, A., 1996: Sea Ice. Part 1. Bulk Salinity Versus Ice Floe Thickness, http://www.dtic.mil/docs/citations/ADA312027.

Lammers, R. B., J. W. Pundsack, and A. I. Shiklomanov, 2007: Variability in river temperature, discharge, and energy flux from the Russian pan-Arctic landmass. *J. Geophys. Res.*, **112**, https://doi.org/10.1029/2006jg000370.

Lindsay, R., and A. Schweiger, 2015: Arctic sea ice thickness loss determined using subsurface, aircraft, and satellite observations. *Cryosphere*, **9,** 269-283, https://doi.org/10.5194/tc-9-269-2015.

Lique, C., and M. Steele, 2012: Where can we find a seasonal cycle of the Atlantic water temperature within the Arctic Basin? *J. Geophys. Res.*, **117**, https://doi.org/10.1029/2011jc007612.

Lique, C., and M. Steele, 2013: Seasonal to decadal variability of Arctic Ocean heat content: A model-based analysis and implications for autonomous observing systems. *J. Geophys. Res.*, **118,** 1673-1695, https://doi.org/10.1002/jgrc.20127.

Lique, C., A. M. Treguier, B. Blanke, and N. Grima, 2010: On the origins of water masses exported along both sides of Greenland: A Lagrangian model analysis. *J. Geophys. Res.*, **115**, https://doi.org/10.1029/2009jc005316.

Lique, C., A. M. Treguier, M. Scheinert, and T. Penduff, 2009: A model-based study of ice and freshwater transport variability along both sides of Greenland. *Climate Dynamics*, **33,** 685-705, https://doi.org/10.1007/s00382-008-0510-7.

MacGilchrist, G. A., A. C. N. Garabato, T. Tsubouchi, S. Bacon, S. Torres-Valdes, and K. Azetsu-Scott, 2014: The Arctic Ocean carbon sink. *Deep Sea Res.*, **86,** 39-55, https://doi.org/10.1016/j.dsr.2014.01.002.

Mauritzen, C., and Coauthors, 2011: Closing the loop - Approaches to monitoring the state of the Arctic Mediterranean during the International Polar Year 2007-2008. *Prog. Oceanogr.*, **90,** 62-89, https://doi.org/10.1016/j.pocean.2011.02.010.

Mayer, M., L. Haimberger, M. Pietschnig, and A. Storto, 2016: Facets of Arctic energy accumulation based on observations and reanalyses 2000-2015. *Geophys. Res. Lett.*, **43,** 10420-10429, https://doi.org/10.1002/2016GL070557.

McPhee, M. G., 2013: Intensification of Geostrophic Currents in the Canada Basin, Arctic Ocean. *J. Climate*, **26,** 3130-3138, https://doi.org/10.1175/Jcli-D-12-00289.1.

McPhee, M. G., A. Proshutinsky, J. H. Morison, M. Steele, and M. B. Alkire, 2009: Rapid change in freshwater content of the Arctic Ocean. *Geophys. Res. Lett.*, **36**, https://doi.org/10.1029/2009gl037525.

National_Geophysical_Data_Center, 2006: 2-minute Gridded Global Relief Data (ETOPO2) v2, National Geophysical Data Center, NOAA, https://doi.org/10.7289/V5J1012Q

Ono, N., 1967: Specific heat and heat of fusion of sea ice. *Physics of Snow and Ice*, H. Oura, Ed., Inst of Low Temp. Sci., Hokkaido, Japan, 599-610.

Overeem, I., and J. P. M. Syvitski, 2010: Shifting Discharge Peaks in Arctic Rivers, 1977-2007. *Geografiska Annaler Series a-Physical Geography*, **92a,** 285-296,

Pemberton, P., J. Nilsson, M. Hieronymus, and H. E. M. Meier, 2015: Arctic Ocean Water Mass Transformation in S-T Coordinates. *J. Phys. Oceanogr.*, **45,** 1025-1050, https://doi.org/10.1175/Jpo-D-14-0197.1.

Polyakov, I. V., and Coauthors, 2011: Fate of Early 2000s Arctic Warm Water Pulse. *Bull. Amer. Meteor. Soc.*, **92,** 561-566, https://doi.org/10.1175/2010bams2921.1.

Polyakov, I. V., and Coauthors, 2017: Greater role for Atlantic inflows on sea-ice loss in the Eurasian Basin of the Arctic Ocean. *Science*, **356,** 285-291, https://doi.org/10.1126/science.aai8204.

Polyakov, I. V., A. V. Pnyushkov, and L. A. Timokhov, 2012: Warming of the Intermediate Atlantic Water of the Arctic Ocean in the 2000s. *J. Climate*, **25,** 8362-8370, https://doi.org/10.1175/Jcli-D-12-00266.1.

Porter, D. F., J. J. Cassano, M. C. Serreze, and D. N. Kindig, 2010: New estimates of the large-scale Arctic atmospheric energy budget. *J. Geophys. Res.*, **115,** D08108, https://doi.org/10.1029/2009jd012653.

Proshutinsky, A., and Coauthors, 2009: Beaufort Gyre freshwater reservoir: State and variability from observations. *J. Geophys. Res.*, **114**, https://doi.org/10.1029/2008jc005104.

R/V Sever cruise report, 2006: *Cruise Report for Russian Navy hydrographic vessel Sever, 23-30 August 2006.*

Rabe, B., and Coauthors, 2014: Arctic Ocean basin liquid freshwater storage trend 1992-2012. *Geophys. Res. Lett.*, **41,** 961-968, https://doi.org/10.1002/2013gl058121.

Renner, A. H. H., and Coauthors, 2014: Evidence of Arctic sea ice thinning from direct observations. *Geophys. Res. Lett.*, **41,** 5029-5036, https://doi.org/10.1002/2014gl060369.

Rudels, B., M. Korhonen, U. Schauer, S. Pisarev, B. Rabe, and A. Wisotzki, 2015: Circulation and transformation of Atlantic water in the Eurasian Basin and the contribution of the Fram Strait inflow branch to the Arctic Ocean heat budget. *Prog. Oceanogr.*, **132,** 128-152, https://doi.org/10.1016/j.pocean.2014.04.003.

Saha, S., and Coauthors, 2010: The Ncep Climate Forecast System Reanalysis. *Bull. Amer. Meteor. Soc.*, **91,** 1015-1057, https://doi.org/10.1175/2010bams3001.1.

Schauer, U., and A. Beszczynska-Moller, 2009: Problems with estimation and interpretation of oceanic heat transport - conceptual remarks for the case of Fram Strait in the Arctic Ocean. *Ocean Science*, **5,** 487-494,

Schauer, U., A. Beszczynska-Möller, W. Walczowski, E. Fahrbach, J. Piechura, and E. Hansen, 2008: Variation of Measured Heat Flow Through the Fram Strait Between 1997 and 2006. *Arctic-Subarctic Ocean Fluxes: Defining the Role of the Northern Seas in Climate*, R. R. Dickson, J. Meincke, and P. Rhines, Eds., Springer, 65-85.

Schauer, U., E. Fahrbach, S. Osterhus, and G. Rohardt, 2004: Arctic warming through the Fram Strait: Oceanic heat transport from 3 years of measurements. *J. Geophys. Res.*, **109**, https://doi.org/10.1029/2003jc001823.

Schlosser, P., and Coauthors, 1999: Pathways and mean residence times of dissolved pollutants in the ocean derived from transient tracers and stable isotopes. *Sci Total Environ*, **238,** 15-30,

Schweiger, A., R. Lindsay, J. L. Zhang, M. Steele, H. Stern, and R. Kwok, 2011: Uncertainty in modeled Arctic sea ice volume. *J. Geophys. Res.*, **116**, https://doi.org/10.1029/2011jc007084.

Serreze, M. C., A. P. Barrett, A. G. Slater, M. Steele, J. Zhang, and K. E. Trenberth, 2007b: The large-scale energy budget of the Arctic. *J. Geophys. Res.*, **112**, https://doi.org/10.1029/2006JD008230.

Serreze, M. C., A. P. Barrett, J. C. Stroeve, D. N. Kindig, and M. M. Holland, 2009: The emergence of surface-based Arctic amplification. *Cryosphere*, **3,** 11-19,

Serreze, M. C., M. M. Holland, and J. Stroeve, 2007a: Perspectives on the Arctic's shrinking sea-ice cover. *Science*, **315,** 1533-1536, https://doi.org/10.1126/science.1139426.

Serreze, M. C., J. Stroeve, A. P. Barrett, and L. N. Boisvert, 2016: Summer atmospheric circulation anomalies over the Arctic Ocean and their influences on September sea ice extent: A cautionary tale. *J. Geophys. Res.*, **121,** 11463-11485, https://doi.org/10.1002/2016jd025161.

Shepherd, A., and Coauthors, 2012: A reconciled estimate of ice-sheet mass balance. *Science*, **338,** 1183-1189, https://doi.org/10.1126/science.1228102.

Shiklomanov, A. I., and R. B. Lammers, 2009: Record Russian river discharge in 2007 and the limits of analysis. *Environ. Res. Lett.*, **4**, https://doi.org/10.1088/1748-9326/4/4/045015.

——, 2014: River ice responses to a warming Arctic-recent evidence from Russian rivers. *Environ. Res. Lett.*, **9,** 035008, https://doi.org/Artn 035008

10.1088/1748-9326/9/3/035008.

Shokr, M., and N. Sinha, 2015: *Sea Ice: Physics and Remote Sensing.* American Geophysical Union, 600 pp.

Smedsrud, L. H., R. Ingvaldsen, J. E. O. Nilsen, and O. Skagseth, 2010: Heat in the Barents Sea: transport, storage, and surface fluxes. *Ocean Science*, **6,** 219-234,

Spreen, G., S. Kern, D. Stammer, and E. Hansen, 2009: Fram Strait sea ice volume export estimated between 2003 and 2008 from satellite data. *Geophys. Res. Lett.*, **36**, https://doi.org/10.1029/2009gl039591.

Talley, L. D., 2003: Shallow, intermediate, and deep overturning components of the global heat budget. *J. Phys. Oceanogr.*, **33,** 530-560, https://doi.org/10.1175/1520-0485.

——, 2008: Freshwater transport estimates and the global overturning circulation: Shallow, deep and throughflow components. *Prog. Oceanogr.*, **78,** 257-303, https://doi.org/10.1016/j.pocean.2008.05.001.

Terashima, T., 1996: Distribution of mesencephalic trigeminal nucleus neurons in the reeler mutant mouse. *Anat Rec*, **244,** 563-571, https://doi.org/10.1002/(SICI)1097-0185.

Timmermans, M. L., and C. Garrett, 2006: Evolution of the deep water in the Canadian basin in the Arctic Ocean. *J. Phys. Oceanogr.*, **36,** 866-874, https://doi.org/10.1175/Jpo2906.1.

Torres-Valdes, S., and Coauthors, 2009: Distribution of dissolved organic nutrients and their effect on export production over the Atlantic Ocean. *Global Biogeochemical Cycles*, **23**, https://doi.org/10.1029/2008gb003389.

Torres-Valdes, S., T. Tsubouchi, E. Davey, I. Yashayaev, and S. Bacon, 2016: Relevance of dissolved organic nutrients for the Arctic Ocean nutrient budget. *Geophys. Res. Lett.*, **43,** 6418-6426, https://doi.org/10.1002/2016gl069245.

Travers, C. S., 2012: Quantifying Sea-Ice Volume Flux using Moored Instrumentation in the Bering Strait. MsC, University of Washington, 77 pp, http://psc.apl.washington.edu/HLD/Bstrait/TraversC2012_MScThesis_BeringStraitIceFlux.pdf.

Uppala, S. M., and Coauthors, 2005: The ERA-40 re-analysis. *Quart. J. Roy. Meteor. Soc.*, **131,** 2961-3012, https://doi.org/10.1256/qj.04.176.

Velicogna, I., 2009: Increasing rates of ice mass loss from the Greenland and Antarctic ice sheets revealed by GRACE. *Geophys. Res. Lett.*, **36**, https://doi.org/10.1029/2009gl040222.

Vihma, T., and Coauthors, 2016: The atmospheric role in the Arctic water cycle: A review on processes, past and future changes, and their impacts. *J. Geophys. Res.*, **121,** 586-620, https://doi.org/10.1002/2015jg003132.

Vinje, T., N. Nordlund, and A. Kvambekk, 1998: Monitoring ice thickness in Fram Strait. *J. Geophys. Res.*, **103,** 10437-10449, https://doi.org/10.1029/97jc03360.

Walin, G., 1982: On the Relation between Sea-Surface Heat-Flow and Thermal Circulation in the Ocean. *Tellus*, **34,** 187-195,

Woodgate, R. A., 2006: Cruise Report for CCGS Sir Wilfrid Laurier 0- SWL2006-01, 1-9 July 2006, http://psc.apl.washington.edu/HLD/SWL2006/UWmooringreportSWL2006July.pdf.

Woodgate, R. A., and K. Aagaard, 2005: Revising the Bering Strait freshwater flux into the Arctic Ocean. *Geophys. Res. Lett.*, **32**, https://doi.org/10.1029/2004gl021747.

Woodgate, R. A., T. Weingartner, and R. Lindsay, 2010: The 2007 Bering Strait oceanic heat flux and anomalous Arctic sea-ice retreat. *Geophys. Res. Lett.*, **37**, https://doi.org/10.1029/2009GL041621.

Wunsch, C., 1996: *The Ocean Circulation Inverse Problem.* Cambridge University press, 442 pp.

Zhang, J. L., and D. A. Rothrock, 2003: Modeling global sea ice with a thickness and enthalpy distribution model in generalized curvilinear coordinates. *Monthly Wea.Review*, **131,** 845-861, https://doi.org/10.1175/1520-0493(2003)131.

TABLE 1: Moored instruments analyzed in this study. Length of data (latest starting date and earliest end date), numbers of moored instruments in each gateway are shown as categorized by type of instruments (MicroCAT, RCMs and ADCP).

| Straits | Start date | End date | MicroCAT | RCM | ADCP | Total |
| --- | --- | --- | --- | --- | --- | --- |
| Davis | 1 Sep 2005 | 30 Sep 2006 | 21 | 9 | 9 | 39 |
| Fram west | 5 Sep 2005 | 3 Sep 2006 | 5 | 12 | 0 | 17 |
| Fram central / east | 26 Aug 2005 | 26 Aug 2006 | 16 | 44 | 0 | 60 |
| BSO | 18 Aug 2004 | 15 Sep 2005 | 0 | 5 | 0 | 5 |
|  | 15 Sep 2005 | 17 Jun 2006 | 0 | 3 | 1 | 4 |
|  | 18 Jun 2006 | 29 Jun 2007 | 0 | 3 | 1 | 4 |
| Bering Russia | 20 Aug 2005 | 24 Aug 2006 | 3 | 1 | 0 | 4 |
| Bering USA | 11 Jul 2005 | 15 Jul 2006 | 2 | 1 | 0 | 3 |
|  | 12 Jul 2006 | 31 Aug 2007 | 1 | 1 | 0 | 2 |
| All Straits | 5 Sep 2005 | 24 Aug 2006 | 46 | 81 | 11 | 138 |

TABLE 2: Definitions of five model layers and corresponding water masses. Five model layers are Surface water (SURF) layer, Upper AW (UAW) layer, AW layer, Intermediate water (IW) layer and Deep water (DW) layers.

| Model layer | Layer group | Upper interface | Lower interface |
|---|---|---|---|
| 1 | SURF | Surface | 27.10 $\sigma_0$ |
| 2 | UAW | 27.10 $\sigma_0$ | 27.50 $\sigma_0$ |
| 3 | AW | 27.50 $\sigma_0$ | 30.28 $\sigma_{0.5}$ |
| 4 | IW | 30.28 $\sigma_{0.5}$ | 32.75 $\sigma_{1.0}$ |
| 5 | DW | 32.75 $\sigma_{1.0}$ | Bottom |
| 6 | Full depth | Surface | Bottom |

TABLE 3: Annual average and standard deviation of net boundary transports of volume (Sv), heat (TW) and FW (mSv). Note that temperature transport (TW-eq) and FW transport (mSv-eq) are sensible to choice of reference values, while heat transport (TW) and FW transport (mSv) are not. The heat transport (TW) and FW transport (mSv) are shown in bold for clarify. Oceanic temperature transport and sea ice sensible heat transport are calculated with reference to the appropriate boundary-mean potential temperature for each month. The annual mean (±1 sd) is 1.01±0.16°C for September 2005 – August 2006. In analogy, FW transports are calculated with reference to the area-weighted mean salinity across the section. That is 34.67±0.02 (annual mean ±1 sd) during September 2005 – August 2006.

|  | Volume transport (Sv) | Temperature / Heat transport (TW-eq / TW) | FW transport (mSv-eq / mSv) |
|---|---|---|---|
| Four main gate ways | | | |
| Davis | -2.1±0.7 | 37±9 | 109±13 |
| Fram | -1.1±1.2 | 63±17 | 79±22 |
| BSO | 2.3±1.2 | 55±28 | 15±12 |
| Bering | 0.7±0.7 | -0±10 | -48±52 |
| Net boundary transports | | | |
| Oceanic | -0.15±0.06 | **154±44** | **155±65** |
| Sea ice | -0.06±0.04 | **22±15** | **48±32** |
| Oceanic plus sea ice | -0.20±0.08 | **175±48** | **204±85** |
| Fram Straits components | | | |
| Belgica | -0.4±0.5 | 3±5 | 11±20 |
| EGC | -6.2±1.2 | 23±7 | 10±8 |
| Middle | -2.0±1.9 | 5±6 | -11±12 |
| WSC | 7.4±1.0 | 32±9 | 69±9 |

TABLE 4: Full depth oceanic heat transport (TW) and its composition of temperature transport (TW-eq) in each water mass in the second column. The velocity driven and temperature driven component of these quantities are also shown in the third and fourth columns. The mean and standard deviation is based on 12 months estimates. For each water masses, only standard deviations are shown because mean values are always close to zero. The heat transports, which are insensitive to choice of reference temperature, are shown in bold for clarify.

|  | Total variability | Velocity driven | Temperature driven |
|---|---|---|---|
| SURF | 47±21 | ±6 | ±17 |
| UAW | 30±10 | ±6 | ±4 |
| AW | 69±18 | ±17 | ±5 |
| IW | 4±3 | ±1 | ±2 |
| DW | 4±6 | ±6 | ±2 |
| Full depth | **154±44** | **2±28** | **-2±25** |

TABLE 5: Same as table 6, but for FW transports (mSv, mSv-eq). The FW transports (mSv), which are insensitive to choice of reference salinity, are shown in bold for clarify.

|  | Total variability | Velocity driven | Salinity driven |
|---|---|---|---|
| SURF | 79±59 | ±56 | ±22 |
| UAW | 31±9 | ±7 | ±6 |
| AW | 57±15 | ±14 | ±3 |
| IW | -9±5 | ±3 | ±3 |
| DW | -3±5 | ±5 | ±2 |
| Full depth | **155±65** | **8±59** | **-0±20** |

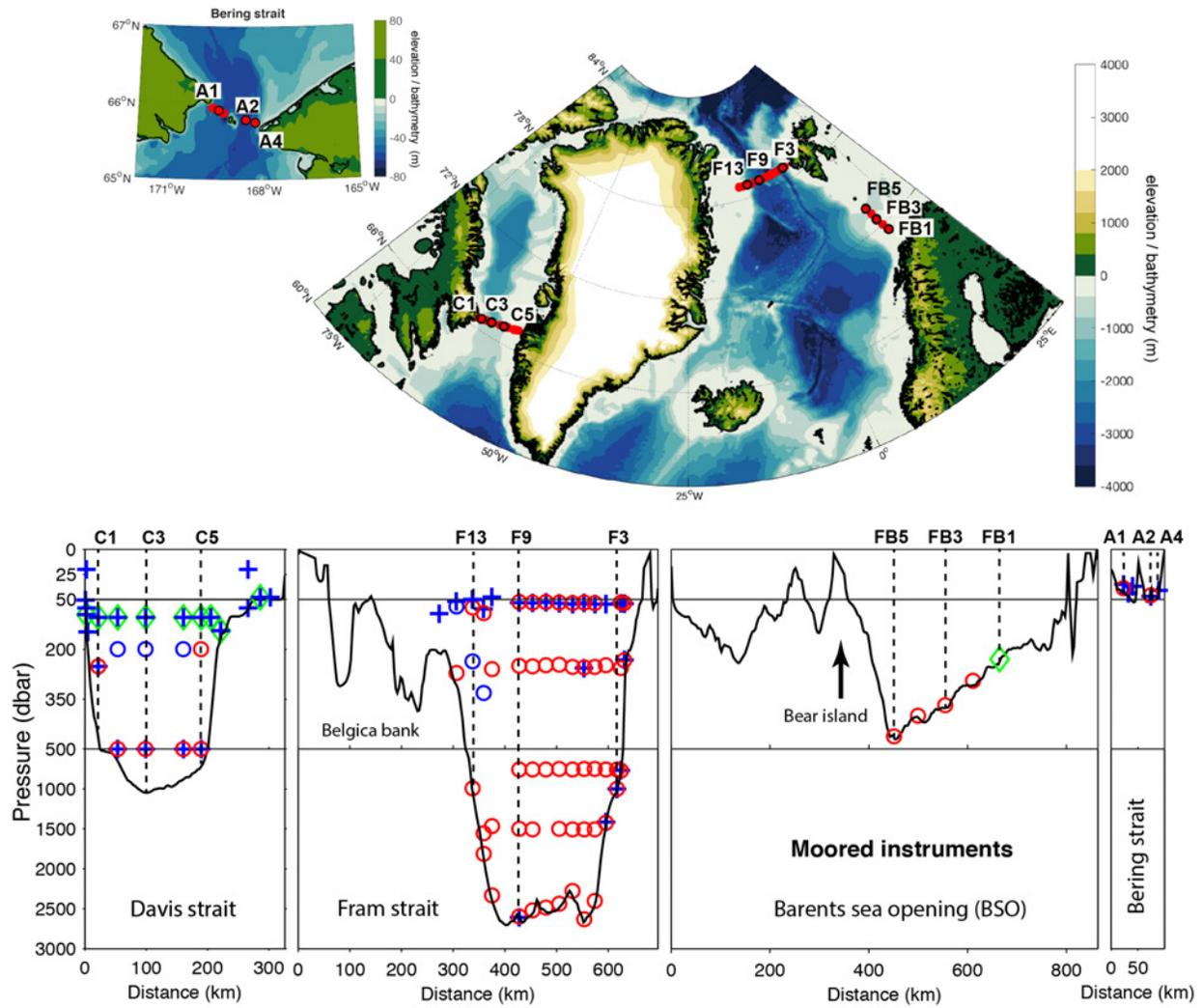

FIG. 1: (Top) Mooring locations in Davis Strait, Fram Strait, the Barents Sea Opening (BSO) and Bering Straits analyzed in this study as red circles. Few mooring sites are labeled and highlighted in black circles. Elevation and Bathymetric data is based on ETOPO2v2 [National Geophysical Data Center, 2006]. (Bottom) Location of 138 mooring instruments across the pan-Arctic boundary section. SeaBird microCAT measuring temperature and salinity are shown as blue crosses. Aanderaa single point current meter measuring temperature and velocity are shown as red circle. When the current meter also measures salinity and the observed salinities are analyzed in this study, the positions of current meters are shown as blue circles. Green diamond shows ADCP (acoustic Doppler Current Profiler), measuring velocity profiles.

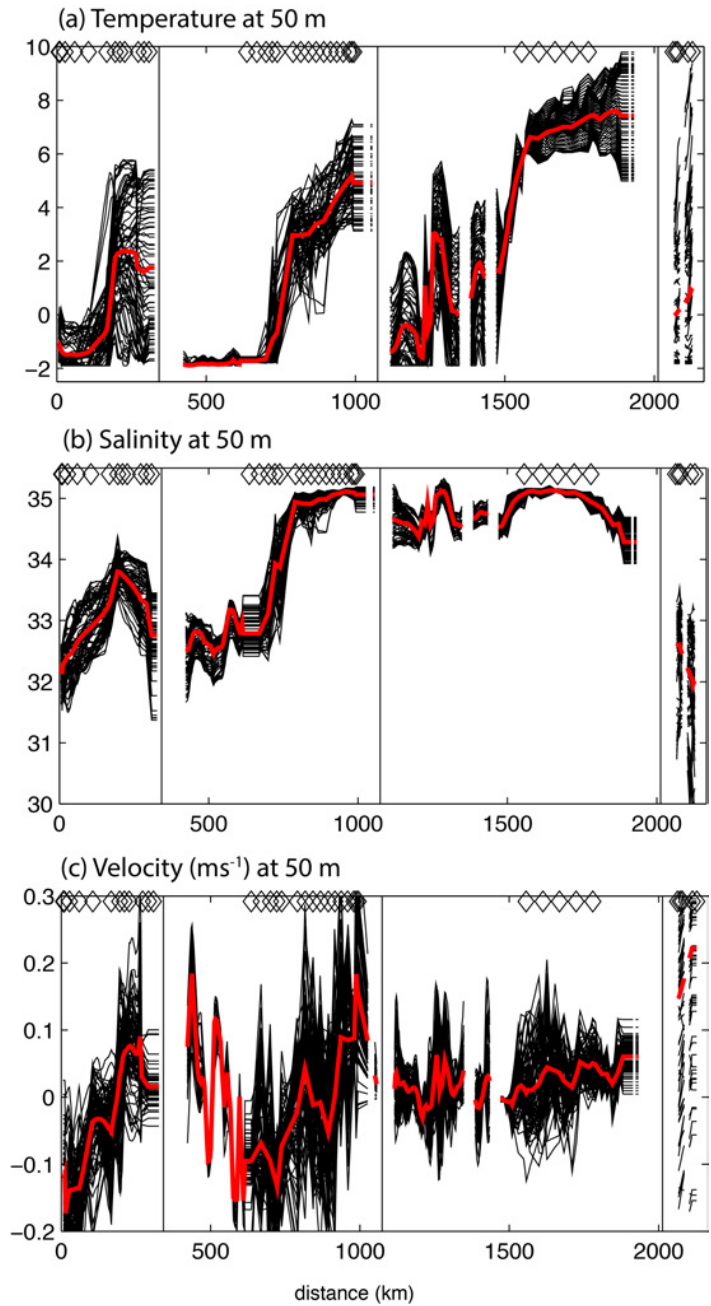

FIG. 2: (a) Constructed temperature variability at 50m depth across the pan-Arctic boundary. Red line shows annual mean value and black lines show individual 5 days temperature. Black diamonds show location of mooring sites. (b) same as (a), but for salinity. (c) same as (a), but for velocity (m s$^{-1}$).

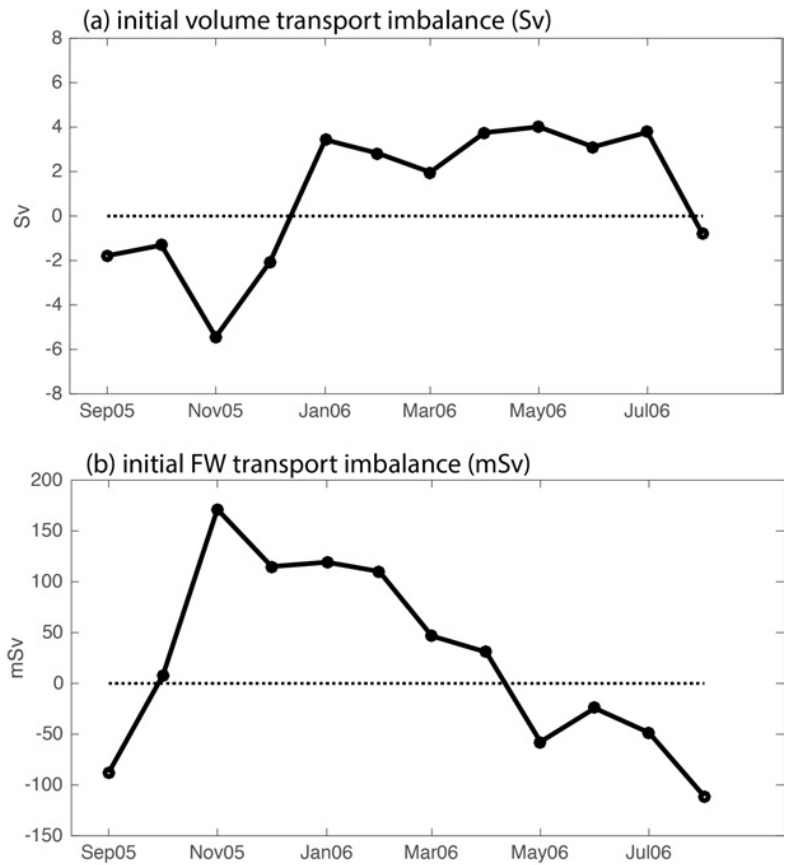

FIG. 3: Time series of initial imbalances for (a) volume transport (Sv) and (b) FW transport (mSv). Zero transports of each property are shown in horizontal dotted lines.

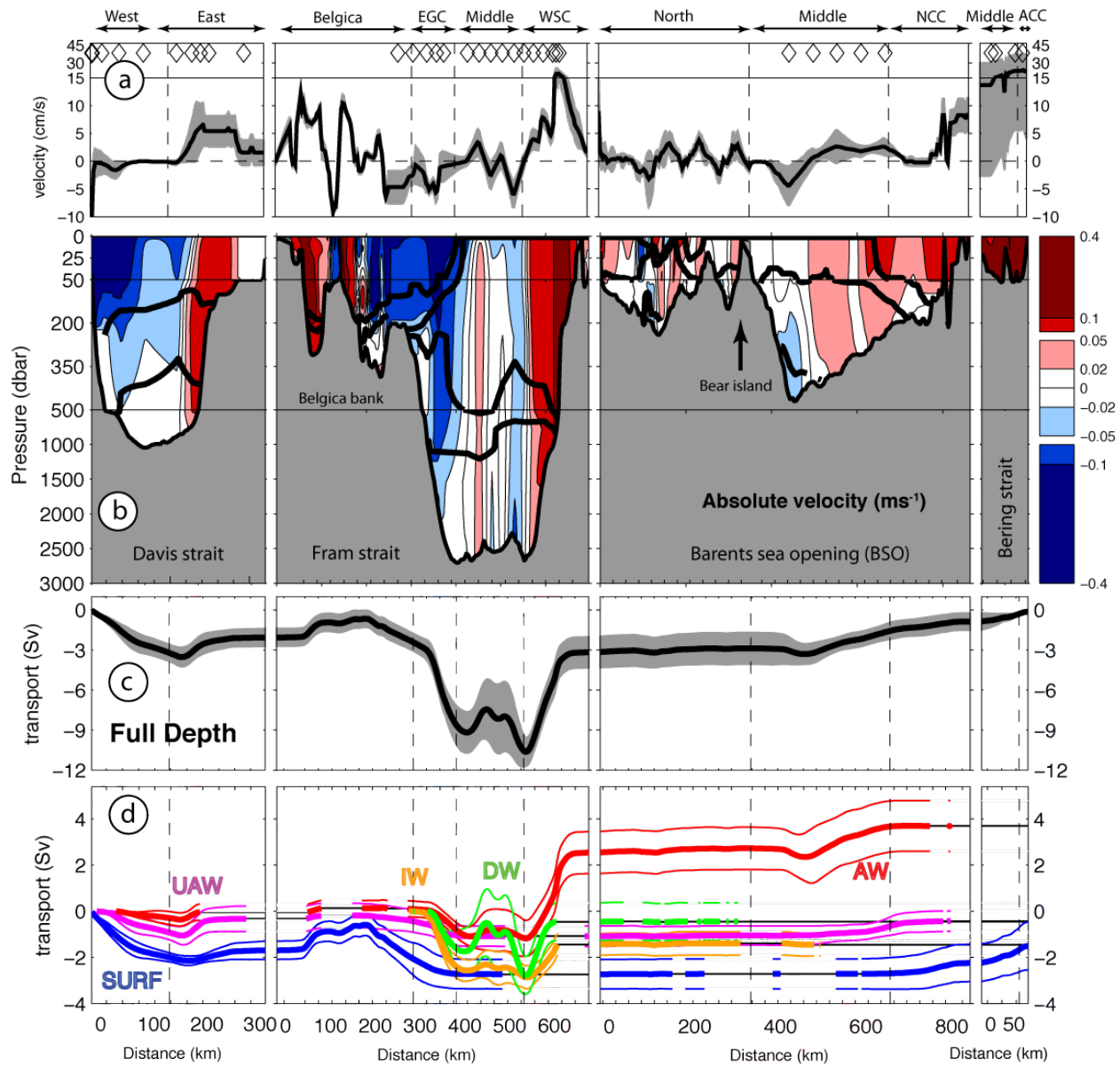

FIG. 4: (a) Inverted annual mean bottom velocities (cm s$^{-1}$; black line) and its standard deviation (grey shade). Mooring locations are shown in diamonds. Note change of vertical scale at 15 cm s$^{-1}$. (b) Volume and salt conserved annual mean velocity field (m s$^{-1}$). Bold black lines show defined water mass boundaries. Red (blue) colors show inflow to (outflow from) the Arctic. (c) inverted annual mean full-depth volume transport (Sv) accumulated around the boundary. Grey shade shows its standard deviation. (d) Annual mean (bold lines) and standard deviation (thin lines) of accumulated volume transport (Sv) for each water mass; where a specific water mass is absent from the section, the accumulated transport is plotted as a black line.

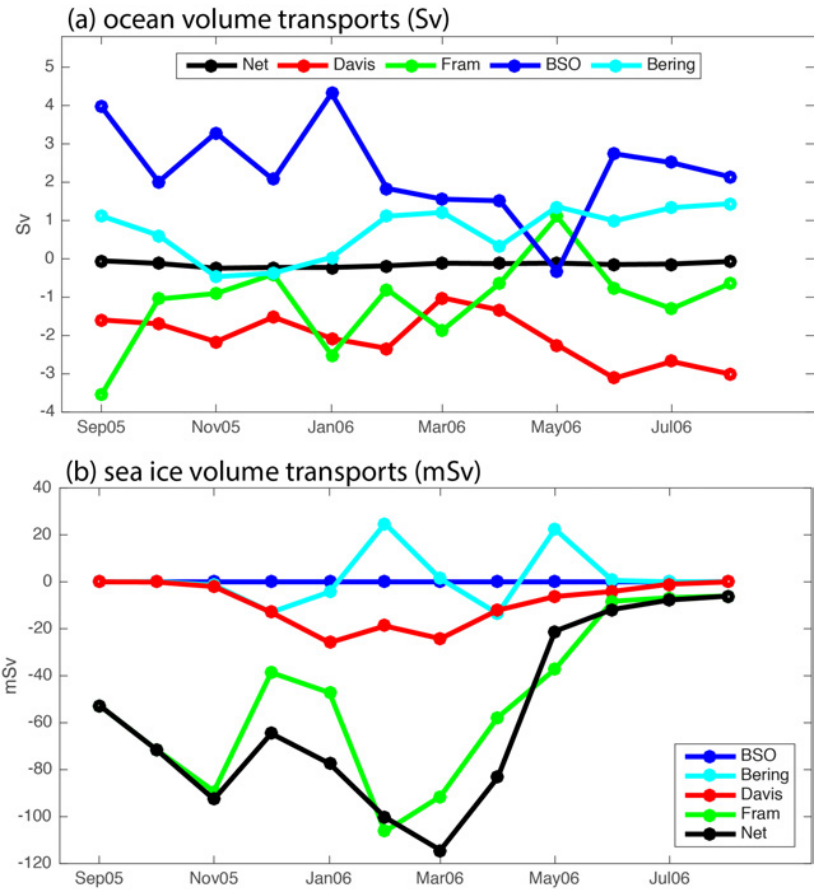

FIG. 5: (a) Oceanic volume transport (Sv) time series in different gateways; Net transport in black and each gateway transports in colours. (b) Sea ice volume transport time series (mSv) in different gateways; Net transport in black and each gateway transports in colours.

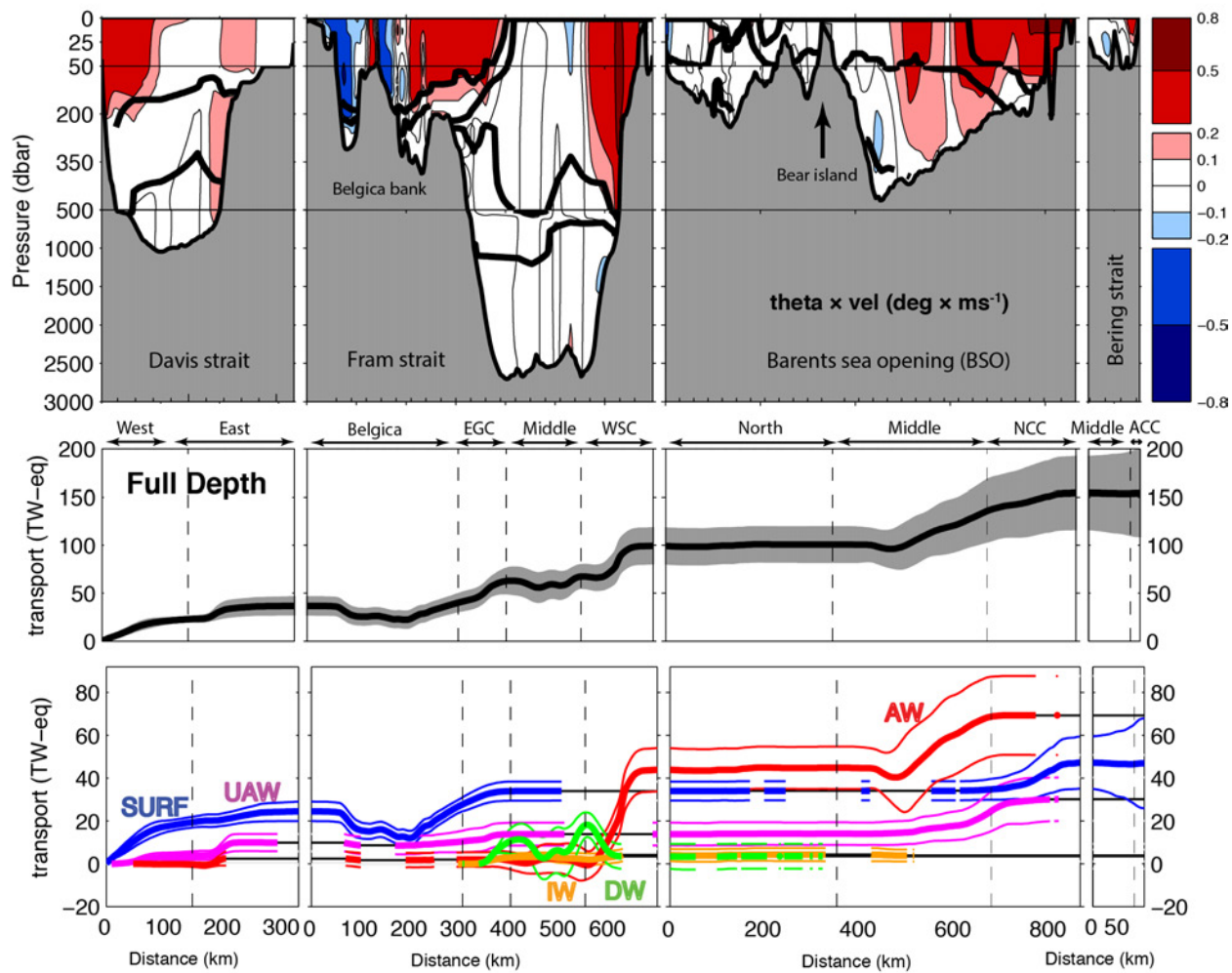

FIG. 6: (top) Annual mean temperature flux (°C×ms$^{-1}$) section calculated from the potential temperature section and final velocity field; bold black lines show defined water mass boundaries, and positive values show temperature entering the Arctic. (middle) annual mean accumulated full depth temperature transport (TW-eq) around the section in black line and its standard deviation in grey shade. (bottom) Annual mean (bold lines) and standard deviation (thin lines) of accumulated temperature transport (TW-eq) of defined water masses; where a specific water mass is absent from the section, the accumulated transport is plotted as a black line.

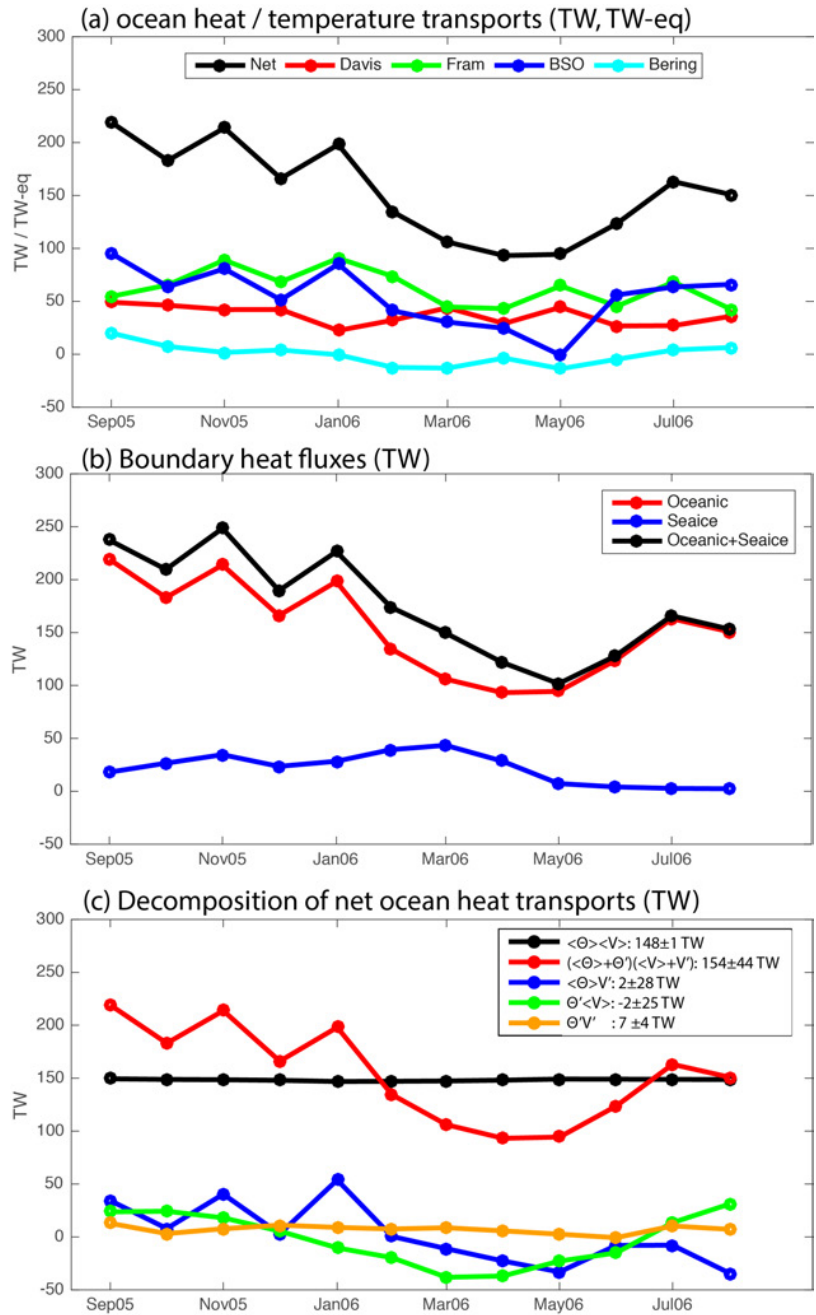

FIG. 7: (a) Oceanic net heat transport (TW) and temperature transport (TW-eq) time series: Net heat transport in black and each gateway temperature transports in colours. (b) Boundary heat flux (TW) time series in black line. Oceanic contribution in red line and sea ice contribution in blue line. (c) Decomposition of net oceanic heat transports (TW) into different components. See detail of the decomposition in main text.

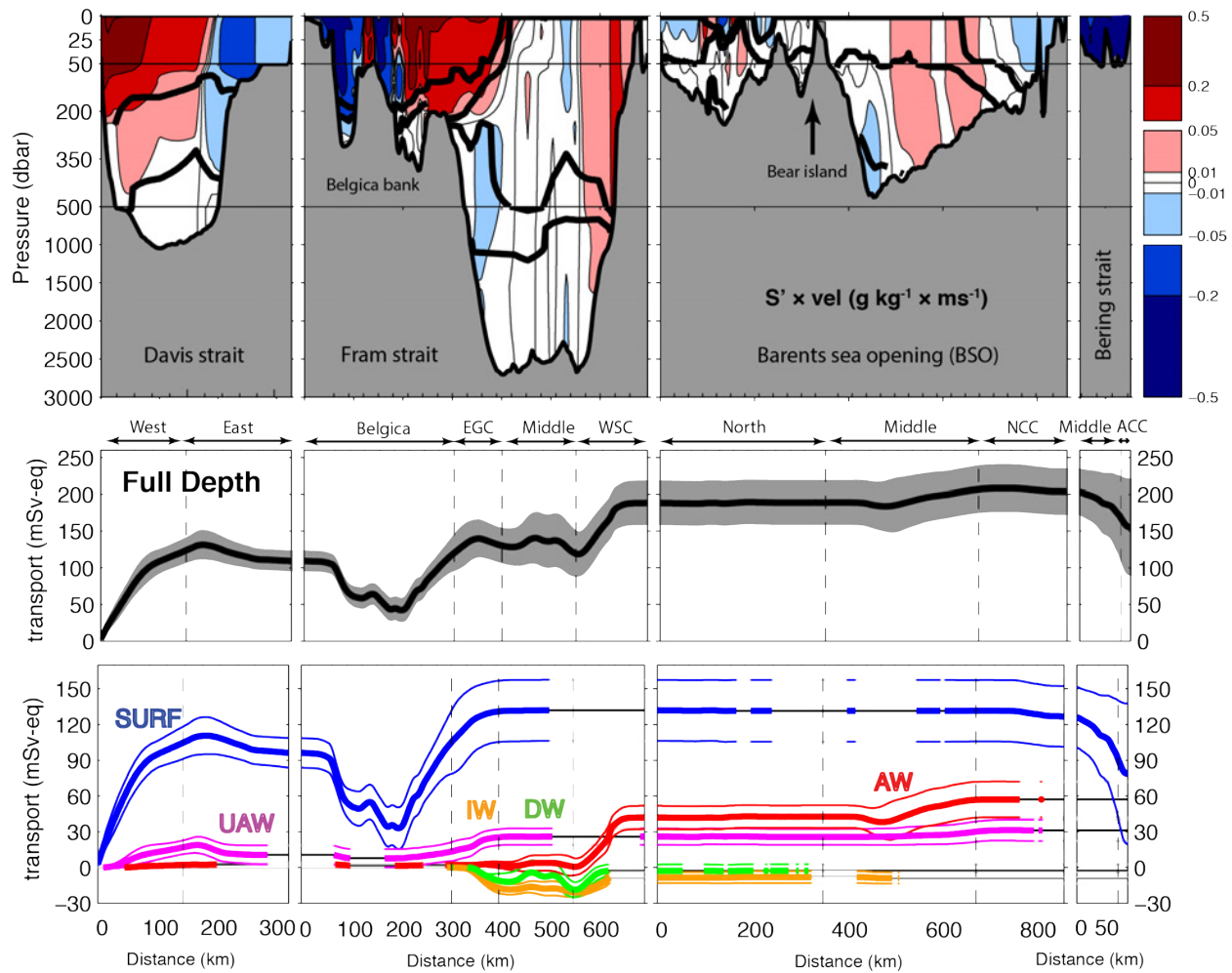

FIG. 8: (top) Annual mean FW flux section (g kg$^{-1}$ × ms$^{-1}$) calculated from the salinity and final velocity fields; bold black lines show defined water mass boundaries. (middle) Anual mean accumulated full depth FW transport (mSv-eq) around the section in black and its standard deviation in grey shade. (bottom) Annual mean (bold lines) and standard deviation (thin lines) of accumulated FW transport (mSv-eq) of the defined water masses. Where a specific water mass is absent from the section, the accumulated transport is plotted as a black line.

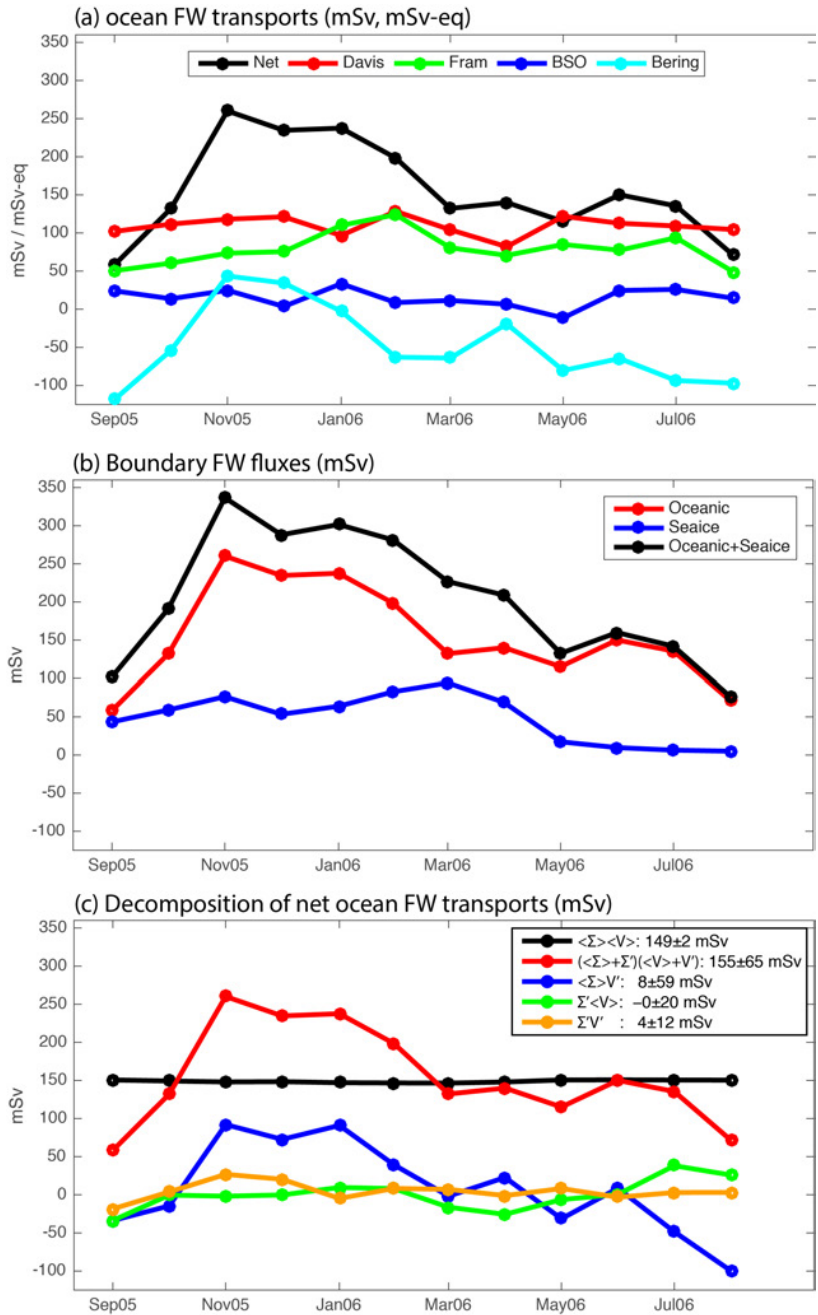

FIG. 9: (a) Oceanic FW transport time series: Net transport (mSv) in black and each gateway transports (mSv-eq) in colours. (b) Boundary FW flux (mSv) time series in black line. Oceanic contribution in red line and sea ice contribution in blue line. (c) Decomposition of net oceanic FW transports (mSv) into different components. See detail of the decomposition in main text.

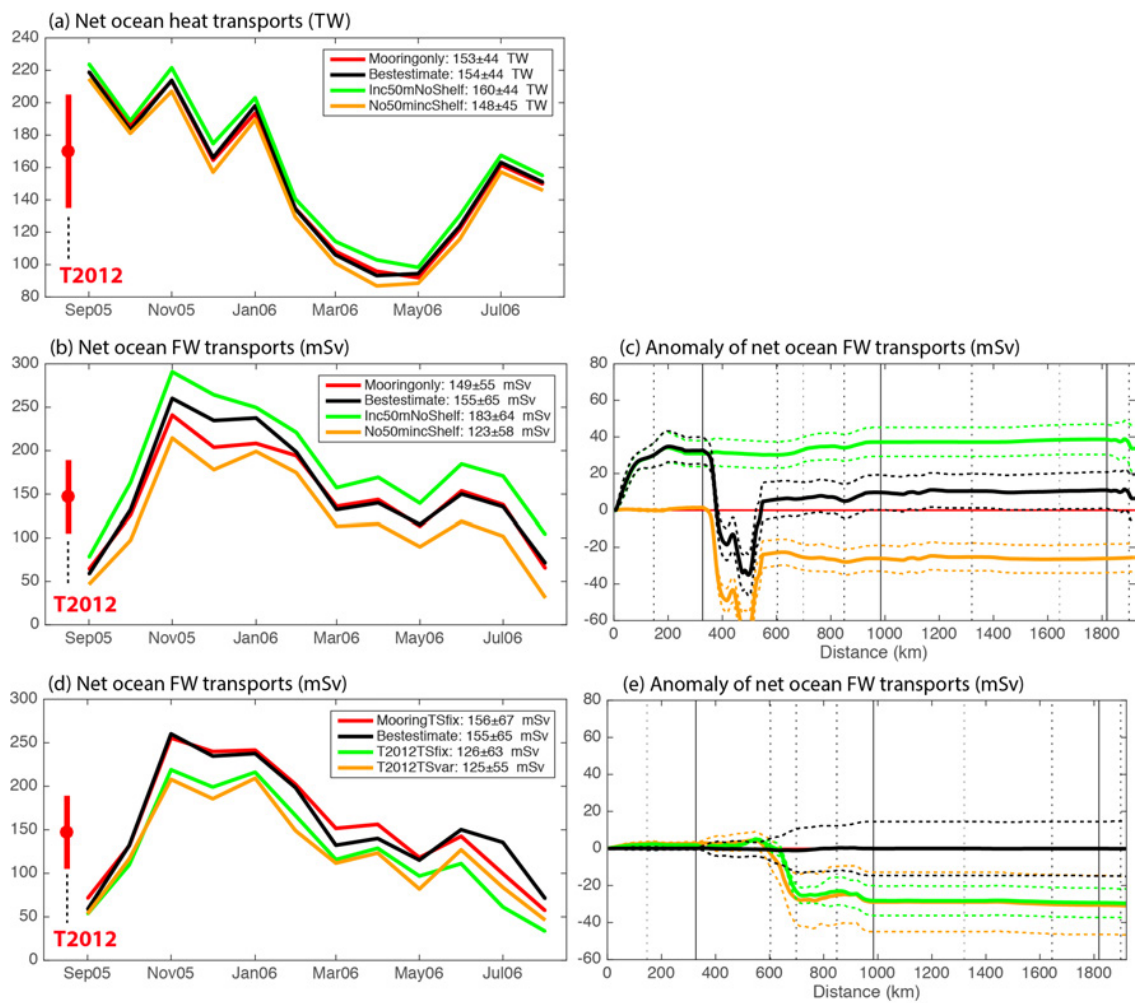

FIG. 10: (a) Net oceanic heat transports (TW) with different set of temperature, salinity and velocity sections to quantify an impact of upper 50m and shelf region variability on the transport estimate. T2012's net oceanic heat transport estimate and its uncertainty estimate are shown at the beginning of the time series in red dot and vertical bar, respectively. (b) same as (a), but for net oceanic FW transport (mSv). (c) Accumulative oceanic FW transports anomaly (mSv) across the Arctic boundary reference to mooring only transport estimate, based on four different FW transport estimates in (b) to quantify impact of upper 50m and shelf region variability on the transport estimate. Solid line shows twelve months mean value and dotted line shows its standard deviation. (d) same as (b), but for quantifying an impact of sparse salinity measurements in Fram Strait on the FW transport estimate (mSv). (e) same as (c), but based on four different FW transport estimates in (d).

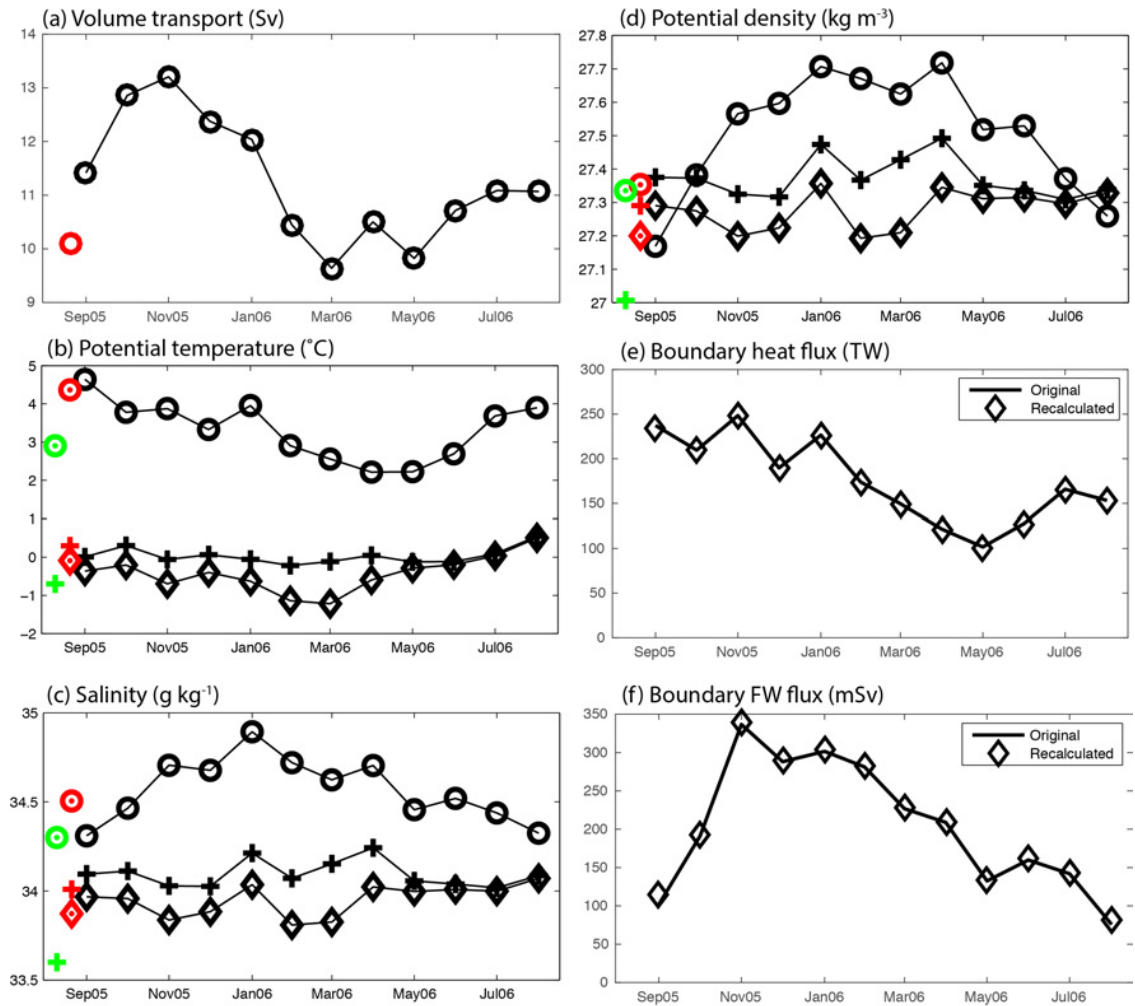

FIG. 11: (a) Volume transport (Sv) in inflow (i.e. sum of water going into the Arctic), associated with the water mass transformation. Updated T2012's estimates are shown at beginning of the time series as red circle. (b) Monthly volume transport weighted potential temperature (˚C) in inflow (circle) and outflow without sea ice (cross), and with sea ice (diamond). T2012's estimates and Pemberton et al [2015]'s estimates are shown at beginning of the time series in red symbols and green symbols, respectively. (c) same as (b), but for salinity. (d) same as (b), but for potential density (kg m$^{-3}$). (e) Original boundary heat flux time series (TW; Fig 8b) in solid line. Boundary heat flux (TW) based on equation (5) is shown in diamond. (f) Original boundary FW flux (mSv; Fig. 10b) in solid line and boundary FW flux (mSv) based on equation (6) in diamond.

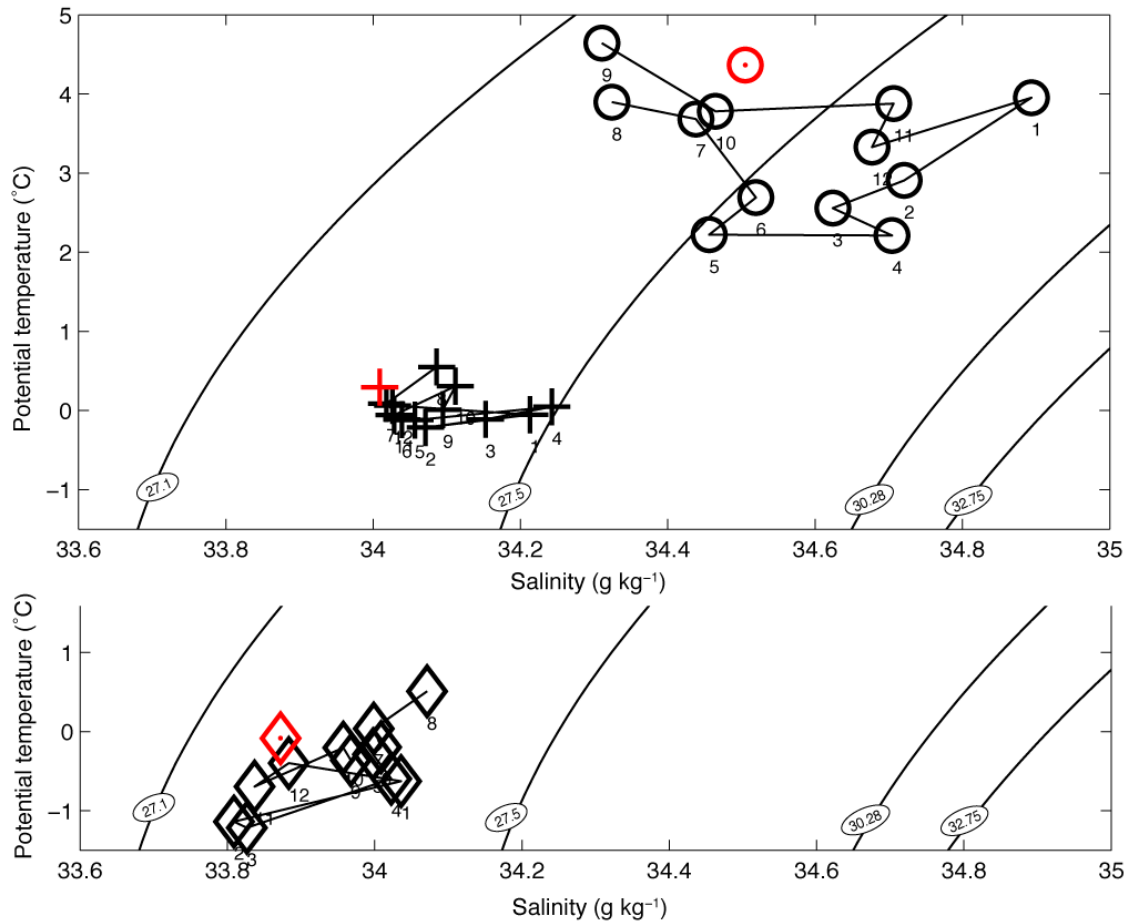

FIG. 12: (Top) Volume transport weighted potential temperature (˚C) and salinity in inflow (circle) and outflow without sea ice (crossed) on θ-S space. Updated estimates of T2012 are shown as red symbols. Number at each plot shows number of months from September 2015 (month 9) towards August 2016 (month 8). Model water mass boundaries (densities) are shown in black. These corresponding densities are 27.1$\sigma_0$, 27.5 $\sigma_0$, 30.28 $\sigma_{0.5}$, 32.75 $\sigma_{1.0}$. (Bottom) same as above, but outflow properties are calculated including sea ice contribution plotted as diamonds. The inflow plots are trimmed to avoid the duplication.

Online supporting Information for

**The Arctic Ocean seasonal cycles of heat and freshwater fluxes: observation-based inverse estimates**

Submitted to Journal of Physical Oceanography in November 2017

Corresponding author: Takamasa Tsubouchi, Geophysical Institute, University of Bergen and Bjerknes Centre for Climate Research, Bergen, Norway
E-mail: takamasa.tsubouchi@uib.no

**Contents**

Support information A: Mooring data and construction of 5 days gridded fields
Support information B: Comparison between September 2005 and T2012
Support information C: Updating main results of T2012

Support information A

**Mooring data and construction of 5 days gridded fields**

*a. Mooring data availability and treatment*

In total, 64 mooring records (42 in temperature, 20 in salinity and 2 in velocity) are not analyzed for following four reasons. All discarded data and corresponding reasons are listed in Table S1-S5. First, short data records (less than 10% of our target period of 1 September 2005 to 31 August 2016) are discarded in order to obtain a seasonal cycle signal with similar mooring array configuration. 7 instrument records (2 in temperature, 3 in salinity and 2 in velocity) from Davis Strait are discarded for this reason (Table S1). Second, temperature and salinity records from RCM or ADCP are discarded when duplicated measurements are available from a nearby MicroCAT (±5 m in depth). The MicroCAT temperature and conductivity measurement accuracies are one order of magnitude better than that of RCM. 27 mooring records (24 in temperature and 3 in salinity) in Davis, Fram and Bering Straits are not analyzed for this reason (Table S1-S3, Table S5). Third, 24 mooring records (13 in temperature and 11 in salinity) in BSO are not analyzed because we analyze the nice repeat CTD sections to generate 5 days temperature and salinity sections in BSO. Finally, some temperature and salinity measurements are discarded when recorded values are not trusted in terns of water mass characteristics in the region. After the individual check, 3 RCM temperature records and 3 RCM salinity records in western Fram Strait are discarded. The corresponding reasons of the elimination are listed in Table S2. In the end, 225 mooring records (86 temperature records, 51 salinity records, and 88 velocity records) are further analyzed.

An 11th order Butterworth filter is first applied to temperature, salinity and velocity data to remove any tidal signal, except Davis Strait because the processed daily data is already de-tided [Curry et al., 2014]. The cutoff frequency (79.16 h) is selected to achieve a reduction of spectrum power by $10^{-10}$ at 24 h periodicity to eliminate the tidal signals around 12 h and 24 h periods. Next, a Gaussian filter with e-folding scale of 5 days and window of 15 days is applied to establish a common time series with 5 day time step. The window size means that if a data gap is shorter than 30 days, the gap is filled with data around the relevant times. If the gap is longer than 30 days, it remains unfilled. Remaining gaps are treated as follows. If the gap is less than half of the whole year, it is filled with the mean value of the remaining data. If the gap is more than half of the whole year, it remains unfilled. This applies to only 3 records: BI3 velocity record at 96 m in Davis Strait and F3 temperature and salinity records at 754 m in Fram Strait (Table S1 and S3). These remaining gaps will be filled by surrounding information as described in next sub-section.

*b. Gridding procedure*

With all instrumental records on a continuous 5 day time base, initial coast-to-coast and sea-bed-to-surface fields of temperature, salinity and cross-section velocity are created for Davis, Bering and Fram Straits as described next. The horizontal grid resolution is 3km, considering the 1/12 NEMO's horizontal resolution. We add variability information above the shallowest moored instruments and across shelf regions using NEMO output. The use of CTD section data in the BSO requires a different procedure, and is described separately.

Although the actual interpolation procedure differs slightly from one gateway to another because of different mooring array configurations, the general procedure is as follows. First, instrumental records are inserted at the relevant position in the grid. For the case of ADCP velocity profiles, single velocity records from a depth bin just above the instruments are used. Second, each gateway section has a set of "key depths", which are nominal standard instrument depths, but in practice, actual instrument depths differ from these key depths, so this step populates the key depths by vertical interpolation, using the nearest instrument location with the second-closest, whether above or below, depending on the case. Third, gaps resulting from missing data at key depths on grid mooring positions are filled using vertical or horizontal interpolation or both, depending on the case. Missing data can arise for the entire period (e.g. F10 mooring in Fram Strait), or via a short data record (this applies to only 3 cases as described above). Choices of gap filling methods are summarised in Table S1-S5. Fourth, horizontal interpolation across key depths creates continuous layers at those depths. Fifth, vertical interpolation at mooring positions creates continuous vertical profiles. We assume no stratification above the shallowest moored instruments at this point. Finally, horizontal interpolation fully populates the grid at all grid depths.

The Davis Strait array has been maintained by UW since 2004. The 14 moorings provide good coverage both of the deep central region of the strait and of the Baffin Island and West Greenland shelf regions [Curry et al. 2011, 2014]. The shallowest instruments are at 25 m (one each side of the strait, over the shelves), and at 100 m in the middle of the strait. The deepest instruments are at 500 m. Key depths for Davis Strait horizontal interpolation are then 0, 25, 50, 100, 200, 500 m. Velocities are extrapolated where necessary, from the deepest measured value down to zero at the bottom. The deepest point across the section has a depth of ~1,000 m. Temperatures and salinities are extrapolated to the bottom as constant values, using the deepest measurements.

Bering Strait measurements have been maintained by UW since 1990 on the eastern side [Roach et al 1995; Woodgate et al 2005], and on the western side under the Russian-American Long-term Census of the Arctic (RUSALCA) project since 2004 [Woodgate et al 2015]. Interpolation in Bering Strait is straightforward, assuming an unstratified water column.

The Fram Strait array has been maintained by NPI (western side) and AWI (central and eastern Fram Strait) since 1997 [de Steur et al. 2009, 2014; Fahrbach et al. 2001, Schauer et al. 2004, 2008, Beszczynska-Möller et al. 2012]. Across the deep channel, 17 moorings generate temperature, velocity and some salinity measurements at 6 depths between 60 m and the bottom. Key depths for Fram Strait horizontal interpolation is 0, 50, 100, 250, 300, 400, 500, 750, 1000, 1500, 2000 and 2500 m. In 2005-2006, the array lacked data from the F10 mooring (2.0°W). Also, the F3 mooring (8.0°E) lacked data from the upper two instruments (60, 250 m) throughout the year, and also from the instrument at 750 m from November 2005 onwards.

Construction of Fram Strait salinity fields is problematic because measurements are sparse, with 16 instruments across the strait around 60 m water depth, 4 around 250 m, and 4 salinity measurements below 750 m. We note that T2012 shows that the part of Fram Strait deeper than 1000 m makes a negligible contribution to net freshwater fluxes; that the dominant contribution is found shallower than 200 m; and that waters between 200–500 m make a modest contribution. Therefore we proceed on the assumption that the available salinity measurements (supplemented by NEMO output) are adequate to capture the variability of the inflowing Atlantic waters and outflowing Polar waters – an assumption that we test through calculations in section 3.5.

Two full-depth salinity profiles are generated, one at F9 (0.8ºW), the other at F4 (7.0ºE). F4 is assigned 5 "nodes" (pseudo-instrument locations) at 64, 250, 766, 1001 and 1415 m. The F4 mooring has salinity measurements at the shallowest and deepest nodes. Salinities at the 766 and 1001 m nodes are populated from the instruments on the adjacent moorings, F2 and F3, respectively. The 250 m node is populated by the average of the measured salinities from the nearest instruments to either side at that depth, on F1 and F6. F9 proceeds similarly, with 5 nodes at 60, 250, 1001, 1415 and 2609 m. The F9 mooring also has salinity measurements at the shallowest and deepest nodes. The 1001 and 1415 m nodes are populated with data from instruments near those depths on F3 and F4, respectively. The 250 m node is filled with a weighted sum of data from the nearest instruments above (F9, 60 m, 50%), to the west (F13, 237 m, 21%) and to the east (F6, 257 m, 29%), where the figures in parentheses show mooring identifier, instrument depth and weight. The horizontal weights (21%, 29%) are based on the horizontal instrument separations. Four upper-ocean salinity profiles are derived from other available salinity measurements at F13 (5ºW, 50–250 m), F12 (4ºW, 300–400 m), F6 (5ºE, 50–300 m), F1 (8.7ºE, 50–300 m). The two full-depth, four upper-ocean salinity profiles and ten salinity measurements around 60 m are then horizontally interpolated to derive complete salinity fields.

BSO observations have been maintained by IMR since 1997 [Ingvaldsen et al. 2001; Smedsrud et al. 2012]. The BSO fields are constructed by combination of 9 repeat, full-depth (surface to sea bed) hydrographic sections between August 2005 to October 2006, and bottom velocity measurements observed at 5 mooring sites between FB1 at 71.5ºN and FB5 at 73.5ºN. Temperature and salinity measurements, and derived geostrophic velocity profiles, are inserted into the section grid. Linear interpolation is applied in time to obtain 5 day fields. Moored velocity measurements are then smoothed and interpolated into 5 day fields. Observed velocity values from the southern (FB1) and northern (FB5) instruments are tapered horizontally and linearly to zero at 50 km outside the observed region. Finally, 5 days absolute velocity fields are obtained by referencing the geostrophic velocities to the measured bottom velocities. FB5 measurements are missing between 15 September 2005 and 17 June 2006, corresponding to 76% of our target period. Since FB5 sits in the outflow of the Bear Island current, it is important to represent its seasonal cycle, and we produce this using velocity measurements from 18 August 2004 – 15 September 2005 and 18 July 2006 – 29 June 2007. Original data are used where available – 18 July 2006 to 31 August 2006.

To this point, we have populated temperature, salinity and velocity fields across the pan-Arctic boundary through Davis, Fram and Bering Straits and the BSO. We next add variability information above the shallowest moored instruments and across shelf regions using NEMO output. To extrapolate upwards from the shallowest instruments to the surface, we use the vertical gradients of temperature, salinity and velocity from NEMO. This information is used above 100 m in the central Davis Strait, and above 50 m in Fram and Bering Straits. No information is added in BSO where we have full-depth hydrographic data. For the shelf regions, we use the NEMO time evolution of temperature, salinity and velocity variability, initialised with values from T2012. The shelf regions comprise Belgica Bank in Fram Strait (west of 9°W), and north of Bear Island in the BSO (north of 74.5°N).

Support information B

**Comparison between September 2005 and T2012**

Fig. S1 show that there are differences in net oceanic heat and FW transports between the September 2005 estimate (the first month based on moorings, hereafter M01) and the T2012 estimate. The M01 net oceanic heat transport estimate is larger than T2012's estimate by 52 TW and its FW transport estimate is lower than T2012's estimate by -82 mSv. These differences must come from either different sampling timing (synopticity) or different sampling resolution. While the M01 estimate is synoptic within the calendar month from 1-30 September, its spatial sampling resolution is mainly limited by the mooring locations. The T2012 estimate is quasi-synoptic (dates as above), but it is based on high spatial resolution CTD measurements. In order to identify causes of the transport discrepancy, two additional cumulative heat and FW transports are calculated. The first calculation combines the temperature and salinity sections from T2012 with the velocity field from M01. This calculation explains deviations in M01 transports due to differences in the velocity fields. The second estimate combines the temperature and salinity sections from M01 with the velocity field from T2012. This calculation explains deviations in M01 transports due to differences in temperature and salinity sections.

Fig. S1a shows the full-depth cumulative volume transports of M01 and T2012. In comparison with T2012, the M01 horizontal volume transport is characterized by stronger inflow in eastern Davis Strait by +1.5 Sv, stronger export in EGC region in Fram Strait by -1.4 Sv, stronger net export in the Fram Strait middle region by -2.1 Sv, and stronger inflow in WSC region in Fram Strait by +1.8 Sv. Horizontal transports in the BSO and Bering Strait are similar.

Fig. S1b shows the anomalous M01 full-depth heat transports compared with T2012. Anomalous temperature transports are seen in Davis Strait (23 TW-eq), the EGC region in Fram Strait (10 TW-eq), the NCC region in the BSO (13 TW-eq), and Bering Strait (7 TW-eq). These explain the excess heat transport of 52 TW. Inspecting the driving factors of the anomalous heat transports, the velocity-driven contribution generally explains the anomalous M01 temperature transports in Davis Strait and Fram Strait. The temperature-driven contribution is mainly responsible for the anomalous transport in BSO and Bering Strait. Overall, the difference in velocity fields increases the M01 net heat transport by 37 TW, while the difference in temperature fields increases the net heat transport by 16 TW.

Fig. S1c shows the anomalous M01 full-depth FW transport compared with T2012. The anomalous FW transports are seen in eastern Davis Strait (-22 mSv-eq), Fram Strait (-9 mSv-eq), the BSO (-8 mSv-eq), and Bering Strait (-46 mSv-eq). Velocity is the primary driver for anomalous M01 FW transports in Davis Strait, Fram Strait, BSO and Russian side of Bering Strait. While salinity change is a driver for the US side of Bering Strait only, it has a significant impact on the FW transport estimate, reducing it by 35 mSv-eq. We note that after CTD measurements were made on 23 August 2005 (analyzed in T2012), the salinity drops by about 2.0 for a month both at A2 and A4 mooring sites on the US side of Bering Strait. Assuming the volume transport in the region to be 0.5 Sv, this salinity change corresponds to 30 mSv-eq FW transport change.

To conclude, net oceanic heat and FW transport discrepancies between M01 and T2012 are primarily due to velocity changes, while temperature and salinity changes in the BSO and Bering Strait have secondary impacts on the transport difference. Returning to the original question of synopticity or spatial resolution, synopticity certainly matters. However, large natural temporal variability in velocity, temperature and salinity fields between the two estimates inhibits the quantification of the impact of spatial sampling resolution.

Support information C

**Updating main results of T2012**

We report here updated results of T2012, such as pan-Arctic volume, heat and fresh water (FW) transport estimates and associated water mass transformation based on quasi-synoptic CTD and mooring measurements in summer 2005. This correction is motivated by following three reasons: (1) Including International Bathymetric Chart of the Arctic Ocean (IBCAO) bathymetry to estimate coast-to-coast, surface to bottom volume, heat and FW transport estimates; (2) Updating constraints, weight settings and uncertainty settings of the box inverse model as described in section 2.3.2 in this paper; (3) Correcting volume transport weighted temperature and salinity values both in inflow and outflow.

First of all, overall corrections are small in general. Volume transport estimates in four main gateways (Davis, Fram, Bering Strait and Barents Sea Opening (BSO)) do not change more than 0.1 Sv. The net oceanic heat and FW transports do not change more than 2 TW and 4 mSv, respectively. Main differences are initial volume transport estimates before the inversion (table S6) and volume transport weighted salinity in outflow of 0.06 g kg$^{-1}$.

From now on, we first describe detail of the three updating points. Then, updated results accompanied with figures and tables are presented.

The first updating point is to include IBCAO bathymetry. In practice, following procedures are performed in the each four gateway. First, in situ temperature and salinity profiles at CTD stations and geostrophic velocity profiles at CTD station pairs are placed in the relevant 3km horizontal grid. This horizontal grid line is same as in this study. Second, if the deepest depth of profile is shallower than the IBCAO topography at given grid point, deepest value is extended towards the bottom. Third, horizontal linear interpolation is applied for every 1dbar to obtain gridded temperature, salinity and geostrophic velocity fields in each gateway. Fourth, bottom velocities measured by current meter at mooring sites are interpolated onto the relevant 3km using linear interpolation. The gridded absolute velocity field is then obtained by combining the gridded geostrophic velocity field and the interpolated bottom velocities. Finally, regarding the extrapolation for outside of CTD profiles for both edges in Fram Strait and southern edge of BSO. For the temperature and salinity profiles, same temperature and salinity profiles are extended towards the coasts. For the geostrophic velocity profile and bottom velocity, they are taped down to zero towards the coast.

The second updating point is to change setting of box inverse model. All constraints, weight setting and uncertainty settings for the box inverse model are same as this study. See section 2.3.2 for the further details.

The third updating point is to correct volume transport weighted temperature and salinity both in inflow and outflow. We find a mistake in a script to calculate these quantities. The mistake is amended.

Fig. S2 shows the updated pan-Arctic coast-to-coast, surface to bottom temperature and salinity sections. The sections look similar to original sections (Fig. 5 in T2012). Major differences are different shape of bathymetry in Davis Strait due to relatively sparse CTD stations in Davis Strait (16 CTD profiles in the section) and inclusion of shelf regions of both sides of Fram Strait and south of BSO, due to the extrapolation for outside of the CTD profiles.

Using the gridded pan-Arctic temperature, salinity and absolute velocity sections, initial volume, salt and temperature transport imbalances are calculated. Full depth initial volume transport estimate, including sea ice export and surface FW flux, is -0.91 Sv deficit, instead of original estimate of -5.22 Sv deficit. That of full depth FW transport imbalance is 3 mSv surpurs, instead of original estimate of 15 mSv deficit. Table S6 provides comparison of initial volume transport estimates between original T2012 estimate and updated estimate. The major discrepancy comes from Fram Strait of 3.57 Sv difference, which stems in +0.46 Sv in Belgica Bank, +0.88 Sv in EGC, +1.24 Sv in middle region, +0.99 Sv in WSC. In terms of the defined layer transport difference, deep water layer volume transport difference is the highest of +1.98 Sv (-0.26 Sv updated from -2.24 Sv), followed by Atlantic water layer volume transport difference of +0.82 Sv (+3.01 Sv updated from +2.19 Sv). This is the effect of including IBCAO bathymetry on the initial transport estimates before inversion.

After running the box-inverse model, volume and salt conserved absolute velocity field is obtained. Fig. S3 shows the updated pan-Arctic absolute velocity sections, accompanied with inverse solution and accumulative volume transports. The bottom velocity corrections are small of $5.4 \times 10^{-4}$ ms$^{-1}$ on average due to the smaller initial full depth volume transport imbalance of -0.91 Sv deficit. The velocity section looks similar to original velocity section (Fig. 9 in T2012). The table S7 summarizes the updated transport estimates along with uncertainty estimate. The volume transport estimates in four main gateways (Davis, Fram, Bering Strait and Barents Sea Opening) do not change more than 0.1 Sv. Although the West Greenland current in eastern side of Davis Strait is more visible in the velocity section, volume transport estimate stays same: 0.8 Sv instead of 0.9 Sv. The biggest volume transport discrepancy appears in WSC region in Fram Strait of 4.4 Sv instead of 3.8 Sv.

The updated net oceanic transport is 168±25 TW instead of 170±23 TW (quasi-synoptic estimate ± uncertainty). Combining sea ice contribution and considering additional source of uncertainty (see detail for T2012), we estimate updated boundary heat flux is 188±36 TW. Comparing table S7 and table 3 in T2012, the temperature transport estimates in each segment of the section are same within range of 6 TW-eq. The biggest discrepancy appears in WSC region of 31 TW-eq instead of 25 TW-eq.

The updated net oceanic FW transport is 143±47 mSv instead of 147±42 mSv. Combining sea ice contribution and considering additional source of uncertainty (see detail for T2012), we estimate updated boundary FW flux is 184±51 mSv. Comparing FW transport values in table S7 and table 3 in T2012, the FW transport estimates in each segment of section are same within range of 11 mSv-eq. The biggest discrepancy appears in Fram Strait of 59 mSv-eq instead of 70 mSv-eq. This comes from -7 mSv-eq discrepancy in Belgica bank, -1 mSv-eq discrepancy in EGC, -4 mSv-eq discrepancy in middle of Fram Strait and +3 mSv-eq discrepancy in WSC.

Fig. S4 shows volumetric θ-S plot and associated volume transport weighted potential temperature and salinity both in inflow and outflow. Original equivalent plot is Fig. 21 in T2012. The volume transport weighted mean properties in inflow are 34.51 in salinity, 4.37°C in potential temperature and 27.36 kg m$^{-3}$ in potential density. For the outflow including sea ice are 33.87 in salinity, -0.11°C in potential temperature and 27.20 kg m$^{-3}$ in potential density. Comparing the updated estimates against original estimates, they are similar to original estimates within range of -0.34°C for potential temperature, within range of 0.06 g kg$^{-1}$ for salinity and within range of 0.06 kg m$^{-3}$ for potential density. The major difference is volume transport weighted salinity in outflow of 33.87, instead of 33.81. This is primary due to a mistake in an original script to calculate these quantities.

TABLE S1: Davis Strait mooring instruments locations, instrument types, observed variable. When data is not analyzed, these variables are bracketed. How to fill the data gap at moored instruments position depends on case by case basis. Subjectively determined interpolation methods are also listed.

| No | Mooring sites | Longitude | Depth (m) | Instrument type | Observed variable | | | Data gap filling method |
|---|---|---|---|---|---|---|---|---|
| 1 | BI1 | 61.2°W | 52 | SBE37 | T | S | | |
| 2 | BI2 | 61.2°W | 20 | SBE37 | T | S | | |
| 3 | | | 76 | SBE37 | $(T^8)$ | $(S^8)$ | | Vertical |
| 4 | BI3 | 61.2°W | 96 | SBE37 | T | S | | |
| 5 | | | | ADCP | | | $V^1$ | Horizontal |
| 6 | BI4 | 61.2°W | 148 | SBE37 | T | S | | |
| 7 | C1 | 60.8°W | 104 | SBE37 | T | S | | |
| 8 | | | | ADCP | | | V | |
| 9 | | | 252 | SBE37 | T | S | | |
| 10 | | | | RCM8 | (T*) | | V | |
| 11 | C2 | 60.1°W | 104 | SBE37 | T | S | | |
| 12 | | | | ADCP | | | V | |
| 13 | | | 200 | RCM8 | T | $S^†$ | V | |
| 14 | | | 500 | SBE37 | T | S | | |
| 15 | | | | RCM8 | (T*) | | V | |
| 16 | C3 | 59.1°W | 104 | SBE37 | T | S | | |
| 17 | | | | ADCP | | | V | |
| 18 | | | 200 | RCM8 | T | $S^†$ | V | |
| 19 | | | 500 | SBE37 | T | S | | |
| 20 | | | | RCM8 | (T*) | | V | |
| 21 | C4 | 57.7°W | 104 | SBE37 | T | S | | |
| 22 | | | | ADCP | | | $(V^8)$ | Replaced by $V_{C4\_200m}$ |
| 23 | | | 200 | RCM8 | T | $S^†$ | V | |
| 24 | | | 500 | SBE37 | T | S | | |
| 25 | | | | RCM8 | (T*) | | V | |
| 26 | C5 | 57.0°W | 104 | SBE37 | T | S | | |
| 27 | | | | ADCP | | | V | |
| 28 | | | 200 | RCM8 | T | $(S^{†8})$ | V | Vertical |
| 29 | | | 500 | SBE37 | T | S | | |
| 30 | | | | RCM8 | (T*) | | V | |
| 31 | C6 | 56.7°W | 104 | SBE37 | $(T^8)$ | $(S^8)$ | | Horizontal |
| 32 | | | | ADCP | | | $(V^8)$ | Horizontal |
| 33 | WG1 | 56.3°W | 144 | SBE37 | T | S | | |

| | | | | | | | |
|---|---|---|---|---|---|---|---|
| 34 | | | | ADCP | | | V |
| 35 | WG2 | 55.3°W | 20 | SBE37 | T | S | |
| 36 | | | 76 | SBE37 | T | S | |
| 37 | WG3 | 54.9°W | 48 | SBE37 | T | S | |
| 38 | | | | ADCP | | | V |
| 39 | WG4 | 54.5°W | 48 | SBE37 | T | S | |

* TS data observed by RCM or ADCP are discarded because there are MiacroCAT observation nearby.
† Conductivity is measured by RCM. The measurement accuracy by MicroCAT is one order magnitude better than RCM.
1. Data exists until December 2015. Data coverage is 22%.
8. Data exists less than 10%

TABLE S2: Same as table S1, but for Fram Strait western mooring instruments.

| No | Mooring sites | Longitude | Depth (m) | Instrument type | Observed variable | | | Data gap filling method |
|----|---------------|-----------|-----------|-----------------|---------|-----|---|------------------------|
| 1  | F17 | 8.0°W | 93   | SBE16 | T       | S          |   |                          |
| 2  | F14 | 6.4°W | 56   | SBE37 | T       | S          |   |                          |
| 3  |     |       | 71   | RCM9  | T       | $S^\dagger$ | V |                          |
| 4  |     |       | 272  | RCM7  | T       |            | V |                          |
| 5  | F13 | 5.0°W | 50   | SBE37 | T       | S          |   |                          |
| 6  |     |       | 75   | RCM7  | (T*)    |            | V |                          |
| 7  |     |       | 237  | RCM7  | T       | $S^\dagger$ | V |                          |
| 8  |     |       | 995  | RCM8  | T       |            | V |                          |
| 9  | F12 | 4.0°W | 79   | SBE37 | T       | S          |   |                          |
| 10 |     |       | 91   | RCM9  | (T*)    | $(S^{\dagger}*)$ | V |                   |
| 11 |     |       | 331  | RCM7  | T       | $S^\dagger$ | V |                          |
| 12 |     |       | 1556 | RCM11 | $(T^1)$ |            | V | Replaced by $T_{F9\_1502m}$ |
| 13 |     |       | 1816 | RCM8  | T       | $(S^{\dagger 2})$ | V | Replaced by pseudo $S_{F9}$ profile |
| 14 | F11 | 3.3°W | 48   | SBE37 | T       | S          |   |                          |
| 15 |     |       | 260  | RCM9  | $(T^3)$ | $(S^{\dagger 4})$ | V | Horizontal           |
| 16 |     |       | 1467 | RCM11 | $(T^1)$ |            | V | Replaced by $T_{F9\_1502m}$ |
| 17 |     |       | 2331 | RCM8  | T       | $(S^{\dagger 5})$ | V | Replaced by pseudo $S_{F9}$ profile |

* TS data observed by RCM or ADCP are discarded because there are MiacroCAT observation nearby
† Conductivity is measured by RCM. The measurement accuracy by MicroCAT is one order magnitude better than RCM.

1. T is below -2°C
2. S is too high around 34.8 compared with summer 2005 CTD section.
3. T is too cold around -1°C compared with summer 2005 CTD section.
4. S is above 38.
5. S is too high around 35.2-35.3 compared with summer 2005 CTD section of 34.8-34.9.

TABLE S3: Same as table S2, but for Fram Strait central / eastern mooring instruments. When whole data does not exist depth and instrument types are also bracketed.

| No | Mooring sites | Longitude | Depth (m) | Instrument type | Observed variable | | Data gap filling method |
|---|---|---|---|---|---|---|---|
| 1 | F10-8 | 2.0°W | (61) | (RCM8) | $(T^9)$ | $(V^9)$ | Horizontal |
| 2 | | | (63) | (SBE37) | $(T^9)$ $(S^9)$ | | Horizontal |
| 3 | | | (253) | (ADCP) | | $(V^9)$ | Horizontal |
| 4 | | | (750) | (RCM11) | $(T^9)$ | $(V^9)$ | Horizontal |
| 5 | | | (1506) | (RCM11) | $(T^9)$ | $(V^9)$ | Horizontal |
| 6 | | | (2652) | (RCM8) | $(T^9)$ | $(V^9)$ | Horizontal |
| 7 | F9-7 | 0.8°W | 58 | RCM7 | $(T^*)$ | V | |
| 8 | | | 60 | SBE37 | T  S | | |
| 9 | | | 250 | RCM11 | T | V | |
| 10 | | | 756 | RCM11 | T | V | |
| 11 | | | 1502 | RCM11 | T | V | |
| 12 | | | 2598 | RCM11 | $(T^*)$ | V | |
| 13 | | | 2609 | SBE16 | T  $S^2$ | | |
| 14 | F16-4 | 0.4°E | 59 | RCM8 | $(T^*)$ | V | |
| 15 | | | 61 | SBE37 | T  S | | |
| 16 | | | 251 | RCM11 | T | V | |
| 17 | | | 757 | RCM11 | T | V | |
| 18 | | | 1503 | RCM11 | T | V | |
| 19 | | | 2519 | RCM11 | T | V | |
| 20 | F15-4 | 1.6°E | 57 | RCM8 | $(T^*)$ | V | |
| 21 | | | 59 | SBE37 | T  S | | |
| 22 | | | 249 | RCM11 | T | V | |
| 23 | | | 755 | RCM11 | T | V | |
| 24 | | | (1501) | (RCM8) | $(T^9)$ | $(V^9)$ | T for horizontal, V for vertical |
| 25 | | | 2487 | RCM11 | T | V | |
| 26 | F8-8 | 2.8°E | 60 | RCM8 | $(T^*)$ | V | |
| 27 | | | 62 | SBE37 | T  S | | |
| 28 | | | 247 | RCM11 | T | V | |
| 29 | | | 753 | RCM7 | T | V | |
| 30 | | | 1499 | RCM8 | T | V | |
| 31 | | | 2435 | RCM11 | T | V | |
| 32 | F7-7 | 4.0°E | 62 | RCM8 | $(T^*)$ | V | |
| 33 | | | 64 | SBE16 | T  S | | |
| 34 | | | 253 | RCM8 | T | V | |
| 35 | | | 759 | RCM7 | T | V | |
| 36 | | | 1503 | RCM11 | T | V | |

| | | | | | | | | |
|---|---|---|---|---|---|---|---|---|
| 37 | | | 2281 | RCM11 | T | | V | |
| 38 | F6-9 | 5.0°E | 59 | RCM7 | (T$^*$) | | V | |
| 39 | | | 61 | SBE16 | T | S | | |
| 40 | | | 255 | RCM11 | (T$^*$) | (S$^{\dagger *}$) | V | |
| 41 | | | 257 | SBE37 | T | S | | |
| 42 | | | 751 | RCM8 | T | | V | |
| 43 | | | 1507 | RCM11 | T | | V | |
| 44 | | | 2633 | RCM11 | T | | V | |
| 45 | F5-8 | 6.0°E | 62 | RCM7 | (T$^*$) | | V | |
| 46 | | | 64 | SBE16 | T | S | | |
| 47 | | | 253 | RCM8 | T | | V | |
| 48 | | | 749 | RCM11 | T | | V | |
| 49 | | | 1505 | RCM8 | T | | V | |
| 50 | | | 2401 | RCM11 | T | | V | |
| 51 | F4-8 | 7.0°E | 64 | SBE37 | T | S | | |
| 52 | | | (93) | (ADCP) | | | (V$^9$) | Replaced by V$_{F4\_249m}$ |
| 53 | | | 249 | RCM11 | T | | V | |
| 54 | | | 755 | RCM11 | T | | V | |
| 55 | | | 1415 | SBE37 | T | S | | |
| 56 | | | 1421 | RCM11 | (T$^*$) | | V | |
| 57 | F3-8 | 8.0°E | (62) | (RCM7) | (T$^9$) | | (V$^9$) | Vertical and horizontal$^3$ |
| 58 | | | (64) | (SBE37) | (T$^9$) | (S$^9$) | | Horizontal |
| 59 | | | (253) | (RCM11) | (T$^9$) | | (V$^9$) | T for vertical, V for vertical and horizontal$^4$ |
| 60 | | | 754 | RCM8 | T$^1$ | | V$^1$ | Horizontal |
| 61 | | | 999 | RCM11 | (T$^*$) | | V | |
| 62 | | | 1001 | SBE 16 | T | S | | |
| 63 | F2-9 | 8.3°E | 60 | RCM7 | (T$^*$) | | V | |
| 64 | | | 62 | SBE37 | T | S | | |
| 65 | | | 256 | RCM11 | T | | V | |
| 65 | | | 766 | SBE37 | T | S | | |
| 66 | | | 772 | RCM8 | (T$^*$) | | V | |
| 67 | F1-8 | 8.7°E | 61 | RCM7 | (T$^*$) | | V | |
| 68 | | | 63 | SBE37 | T | S | | |
| 70 | | | 232 | SBE37 | T | S | | |
| 71 | | | 233 | RCM8 | (T$^*$) | | V | |

$^*$ TS data observed by RCM or ADCP are discarded because there are MiacroCAT observation nearby
† Conductivity is measured by RCM. The measurement accuracy by MicroCAT is one order magnitude better than RCM.

1. Data exists until November 2005. Data coverage is 18%.

2. S has a linear trend of 0.09, from 34.91 to 34.82 over 373 days.
3. $V_{F3\_62m} = 0.5*V_{F3\_999m} + 0.25*(V_{F4\_249m} + V_{F2\_60m})$
4. $V_{F3\_253m} = 0.5*V_{F3\_999m} + 0.25*(V_{F4\_249m} + V_{F2\_256m})$
9. Whole data does not exist due to instruments lost during recovery, failure of recording etc.

TABLE S4: Same as table S3, but for Barents Sea Opening mooring instruments. 3 different observation periods of data are used to cover our target period of 1 September 2005 to 31 August 2006.

| No | Mooring sites | Latitude (°N) | Depth (m) | Instrument type | Observed variable | | | Data gap filling method |
|----|---------------|---------------|-----------|-----------------|-------------------|---|---|-------------------------|
| Moored instruments during 18 August 2004 – 15 September 2005, covering 4.1% of our target period of 1 September 2005 to 31 August 2006 ||||||||||
| 1  | FB5 | 73.5°N | 469 | RCM7 | $(T^*)$ | $(S^{\dagger *})$ | V | |
| 2  | FB4 | 73.0°N | 400 | RCM7 | $(T^*)$ | $(S^{\dagger *})$ | V | |
| 3  | FB3 | 72.5°N | 376 | RCM7 | $(T^*)$ | $(S^{\dagger *})$ | V | |
| 4  | FB2 | 72.0°N | 297 | RCM7 | $(T^*)$ | $(S^{\dagger *})$ | V | |
| 5  | FB1 | 71.5°N | 216 | RCM7 | $(T^*)$ | $(S^{\dagger *})$ | V | |
| Moored instruments during 15 September 2005 – 17 June 2005, covering 75.6% of our target period of 1 September 2005 to 31 August 2006 ||||||||||
| 6  | FB5 | 73.5°N | (469) | (RCM7) | $(T^9)$ | $(S^{\dagger 9})$ | $(V^9)$ | $V_{F5\_469}$ from different years |
| 7  | FB4 | 73.0°N | 400 | RCM7 | $(T^*)$ | $(S^{\dagger *})$ | V | |
| 8  | FB3 | 72.5°N | 368 | RCM7 | $(T^*)$ | $(S^{\dagger *})$ | V | |
| 9  | FB2 | 72.0°N | 294 | RCM7 | $(T^*)$ | $(S^{\dagger *})$ | V | |
| 10 | FB1 | 71.5°N | 231 | ADCP | $(T^*)$ | | V | |
| Moored instruments during 18 June 2005 – 29 June 2006, covering 20.3% of our target period of 1 September 2005 to 31 August 2006 ||||||||||
| 11 | FB5 | 73.5°N | 461 | RCM7 | $(T^*)$ | $(S^{\dagger *})$ | V | |
| 12 | FB4 | 73.0°N | 402 | RCM7 | $(T^*)$ | $(S^{\dagger *})$ | V | |
| 13 | FB3 | 72.5°N | 375 | RCM7 | $(T^*)$ | $(S^{\dagger *})$ | V | |
| 14 | FB2 | 72.0°N | (297) | (RCM7) | $(T^9)$ | $(S^{\dagger 9})$ | $(V^9)$ | |
| 15 | FB1 | 71.5°N | 231 | ADCP | $(T^*)$ | | V | |

$^*$ TS data observed by RCM or ADCP are not analyzed because we use repeat CTD sections.
† Conductivity is measured by RCM. The measurement accuracy by MicroCAT is one order magnitude better than RCM.
9 Whole data does not exist due to instruments lost during recovery, failure of recording etc.

TABLE S5: Same as table S4, but for Bering Strait mooring instruments. Russian side and US side of moored instruments are listed separately. For US side, 2 different observation periods of data are used to cover our target period of 1 September 2005 to 31 August 2006.

| No | Mooring sites | Longitude | Depth (m) | Instrument type | Observed variable | | | Data gap filling method |
|---|---|---|---|---|---|---|---|---|
| Russian side moored instruments during 20 August 2005 – 24 August 2006, covering 96% of our target period of 1 September 2005 to 31 August 2006 | | | | | | | | |
| 1 | A1W-05 | 169.6°W | 35 | SBE16 | T | S | | |
| 2 | | | 39 | RCM9 | (T$^*$) | (S$^{\dagger *}$) | V | |
| 3 | A1-05 | 169.4°W | 44 | SBE37 | T | S | | |
| 4 | A1E-05 | 169.3°W | 37 | SBE37 | T | S | | |
| US side moored instruments during 11 July 2005 – 15 July 2006, covering 87% of our target period of 1 September 2005 to 31 August 2006 | | | | | | | | |
| 5 | A2-05 | 168.6°W | 45 | RCM7 | (T$^*$) | | V | |
| 6 | | | 46 | SBE16 | T | S | | |
| 7 | A4-05 | 168.3°W | 41 | SBE37 | T | S | | |
| US side moored instruments during 12 July 2006 – 31 August 2007, covering 13% of our target period of 1 September 2005 to 31 August 2006 | | | | | | | | |
| 8 | A2-06 | 168.6°W | 45 | RCM9 | (T$^*$) | | V | |
| 9 | | | 46 | SBE16 | T | S | | |

$^*$ TS data observed by RCM or ADCP are discarded because there are MiacroCAT observation nearby
† Conductivity is measured by RCM. The measurement accuracy by MicroCAT is one order magnitude better than RCM.

TABLE S6: Comparison of initial volume transport estimates (Sv) between published T2012 values and updated estimate including IBCAO bathymetry. The differences between the two (updated – original) are also shown.

|  | Original initial volume transport (Sv) | Updated initial volume transport (Sv) | Difference Updated – original |
|---|---|---|---|
| **Four main gates** | | | |
| Davis | -3.53 | -3.28 | +0.25 |
| Fram | -5.98 | -2.41 | +3.57 |
| Barents | +3.19 | +3.63 | +0.44 |
| Bering | +0.98 | +1.01 | +0.03 |
| **Net boundary transports** | | | |
| Oceanic | -5.35 | -1.04 | +4.31 |
| Sea ice | -0.05 | -0.05 | ±0.00 |
| Oceanic, sea ice plus surface FW flux | -5.22 | -0.91 | +4.31 |
| **Davis Strait components** | | | |
| West | -4.22 | -4.10 | +0.12 |
| East | +0.68 | +0.82 | +0.14 |
| **Fram Straits components** | | | |
| Belgica | -0.96 | -0.50 | +0.46 |
| EGC | -6.47 | -5.59 | +0.88 |
| Middle | -1.81 | -0.57 | +1.24 |
| WSC | +3.26 | +4.25 | +0.99 |
| **BSO components** | | | |
| North | +0.03 | +0.15 | +0.12 |
| Middle | +2.44 | +2.65 | +0.21 |
| NCC | +0.71 | +0.84 | +0.13 |
| **Bering Strait components** | | | |
| Main | +0.82 | +0.78 | -0.04 |
| ACC | +0.16 | -0.23 | +0.07 |

TABLE S7: Summary of updated transport estimates of volume, temperature, heat and FW, with uncertainty estimate. Transport uncertainty is calculated based on a posteriori error, shown as ± *uncertainty* in italic form. The original table is table 3 in T2012.

|  | Volume transport (Sv) | Temperature / Heat transport (TW-eq / TW) | FW transport (mSv-eq/ mSv) |
|---|---|---|---|
| **Four main gates** | | | |
| Davis | -3.2±*0.9* | 26±*4* | 124±*18* |
| Fram | -1.6±*4.8* | 43±*20* | 59±*43* |
| Barents | 3.7±*0.9* | 87±*15* | 32±*10* |
| Bering | 1.0±*0.3* | 13±*3* | -73±*17* |
| **Net boundary transports** | | | |
| Oceanic | -0.14±*4.93* | 168±*25* | 143±*47* |
| Sea ice | -0.05±*0.01* | 20±*3* | 41±*8* |
| Oceanic plus sea ice | -0.19±*4.93* | 188±*25* | 184±*48* |
| **Davis Strait components** | | | |
| West | -4.1±*0.7* | 18±*2* | 144±*16* |
| East | 0.8±*0.6* | 8±*4* | -20±*9* |
| **Fram Straits components** | | | |
| Belgica | -0.5±*0.7* | 5±*6* | 16±*25* |
| EGC | -5.4±*2.6* | 17±*9* | 2±*19* |
| Middle | -0.1±*3.6* | -11±*15* | 1±*26* |
| WSC | 4.4±*1.6* | 31±*7* | 41±*14* |
| **BSO components** | | | |
| North | 0.2±*0.4* | 2±*3* | 0±*3* |
| Middle | 2.7±*0.7* | 60±*11* | 36±*9* |
| NCC | 0.9±*0.4* | 25±*10* | -4±*3* |
| **Bering Strait components** | | | |
| Main | 0.8±*0.2* | 8±*2* | -51±*15* |
| ACC | 0.2±*0.1* | 5±*2* | -22±*8* |

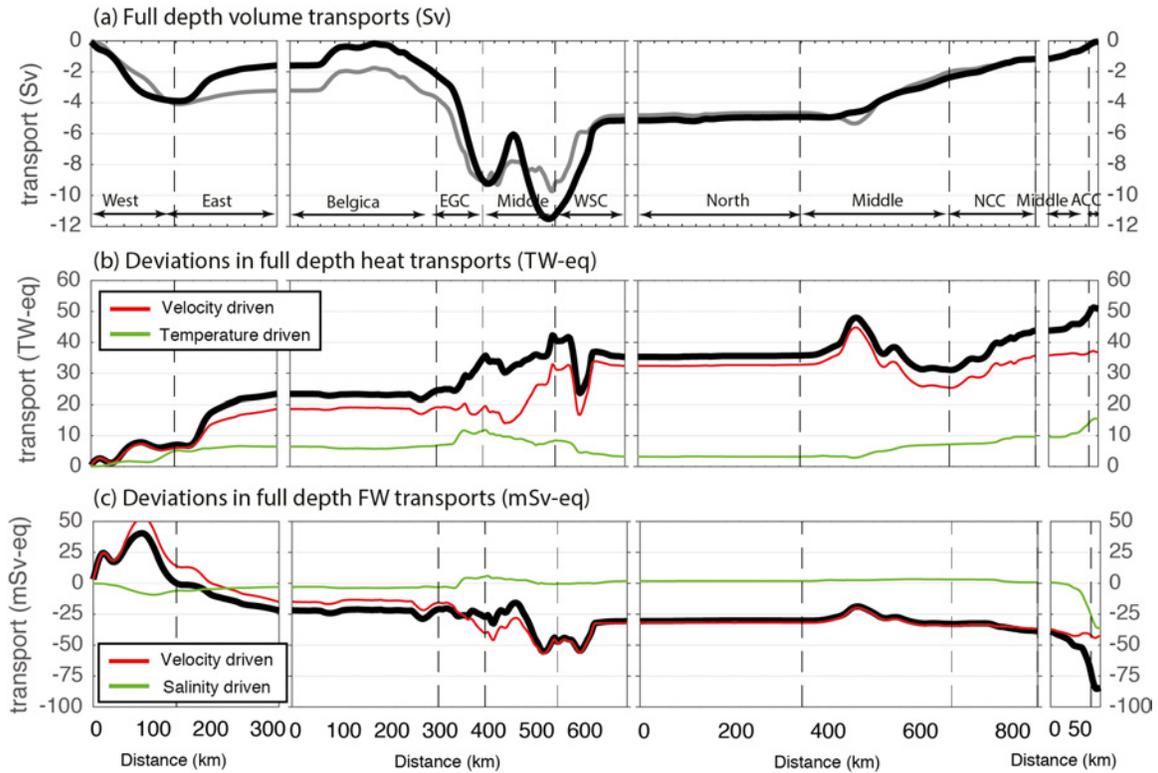

FIG. S1: (a) Accumulative full depth volume transport (Sv) from September 2005 (the first month based on mooring, hereafter M01) in black line and from T2012 in grey line. (b) Deviation in M01 full depth oceanic heat transport (TW-eq) reference to T2012 in black. Velocity driven contribution is shown in red and temperature driven contribution in green. (c) Deviation in M01 full depth oceanic FW transport (mSv-eq) reference to T2012 in black. Velocity driven contribution is shown in red and salinity driven contribution in green.

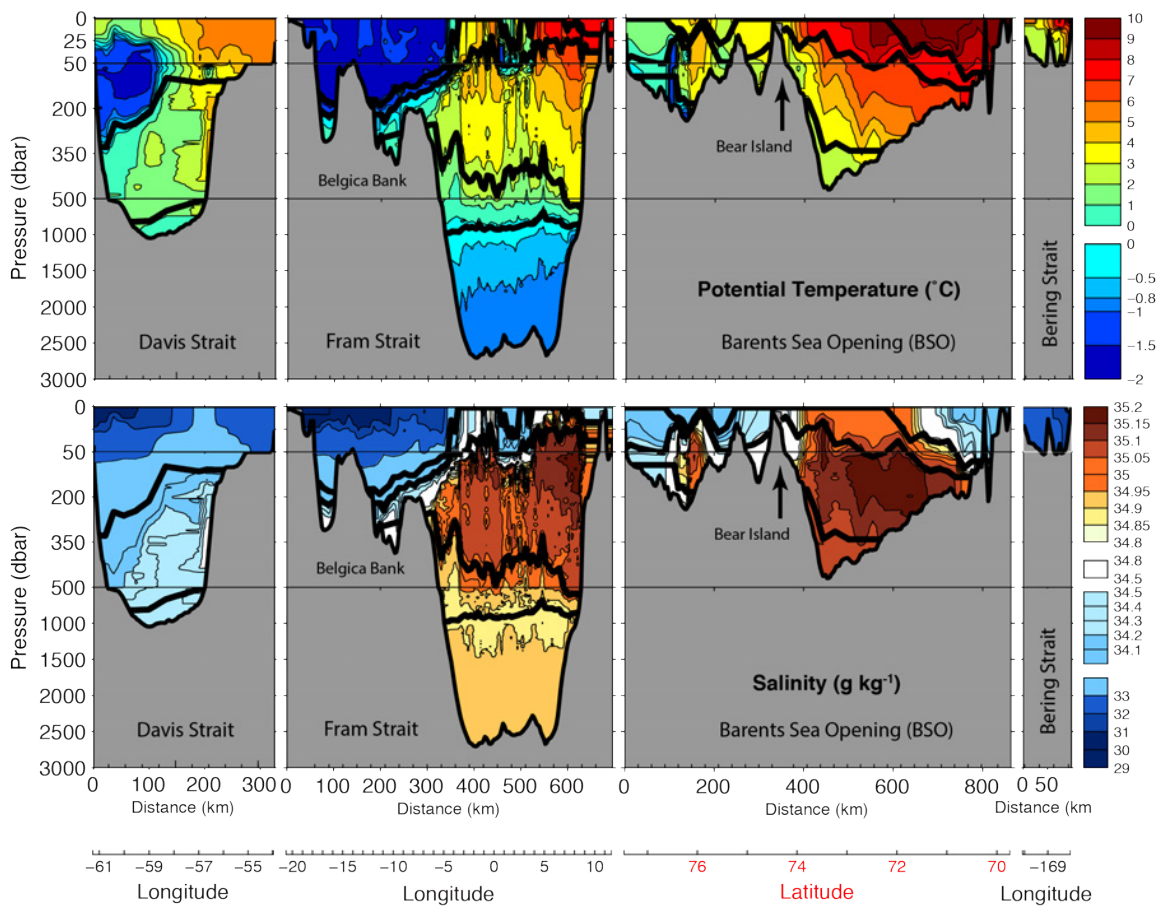

FIG. S2: (Top) Updated pan-Arctic potential temperature section and (bottom) salinity section along Davis and Fram Straits, the BSO and Bering Strait; bold black lines show four defined water mass boundaries; the color bar scale is nonlinear. The pressure axis is expanded between 0-50 dbar and 50-500 dbar. The original plot is Fig. 5 in T2012.

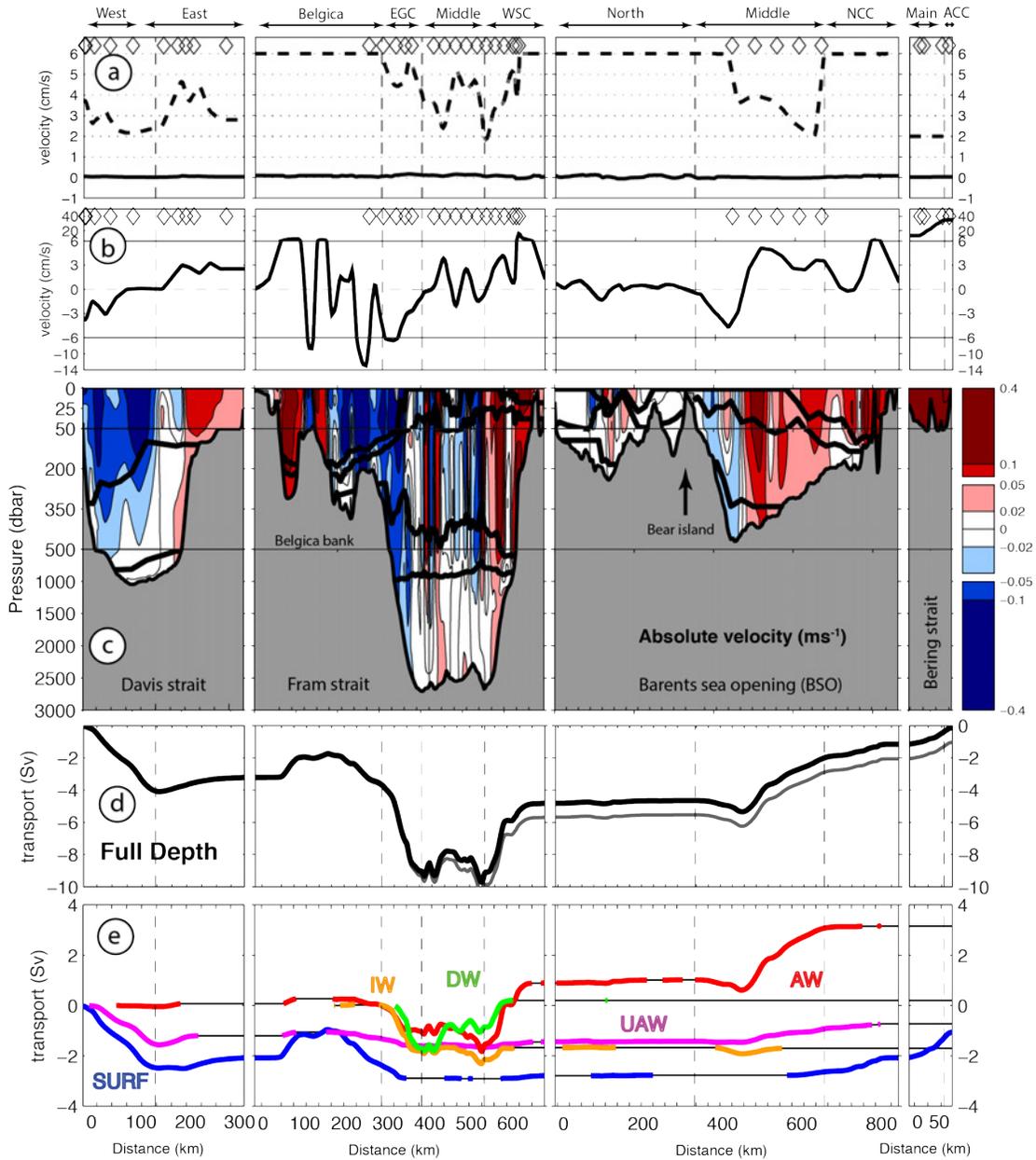

FIG. S3: (a) Standard solution for reference velocities (black solid line), defined *a priori* error for reference velocities (gray dotted line), *a posteriori* error for reference velocities (black dotted line) along the section; mooring locations are shown as diamonds; (b) initial (grey) and final (black) bottom velocities (note change of vertical scale at ±6 cm s$^{-1}$); mooring locations are shown as diamonds; (c) final velocity section (m s$^{-1}$); bold black lines show defined water mass boundaries, red (blue) colours show inflow to (outflow from) the Arctic; (d) initial (gray) and final (black) full-depth volume transport

-24-

(Sv) accumulated around the boundary; (e) cumulative volume transport for each water mass; where a specific water mass is absent from the section, the cumulative transport is plotted as a black line. The original plot is Fig. 9 in T2012.

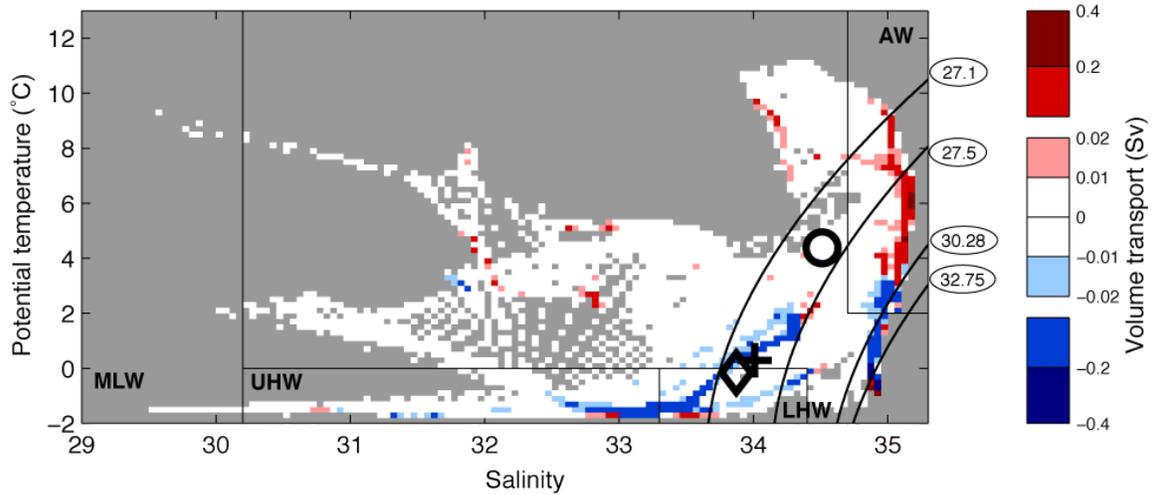

FIG. S4: Volumetric θ-S plot gridded with $\delta\theta = 0.2$°C and $\delta S = 0.05$. Model water mass boundaries (densities) are shown in black. These corresponding densities are 27.1 $\sigma_0$, 27.5 $\sigma_0$, 30.28 $\sigma_{0.5}$, 32.75 $\sigma_{1.0}$. Net transport per θ-S grid box is shaded with red for inflow and blue for outflow (Sv). Grey shading indicates no data. The transport-weighted mean properties of the inflow (bold circle) are: salinity 34.51, potential temperature 4.37°C, density ($\sigma_0$) 27.36 kg m$^{-3}$; for the outflow (bold diamond), including sea ice, they are 33.87, -0.16°C and 27.20 kg m$^{-3}$. Bold cross shows the outflow potential temperature and salinity without sea ice contribution. Some conventional water masses in the central Arctic (MLW, UHW, LHW, AW) are shown based on Aksenov et al. [2010; table 2]. The original plot is Fig. 21 in T2012.